\documentclass[Style/sn-mathphys]{Style/sn-jnl}

\graphicspath{{Figures/}}
\usepackage{amsmath}
\DeclareMathOperator*{\argmax}{arg\,max}
\DeclareMathOperator*{\argmin}{arg\,min}

\usepackage{etoolbox,xspace}

\newcommand{\definetextttcmd}[1]{\csdef{#1}{\texttt{#1}\xspace}}
\forcsvlist{\definetextttcmd}{PORTALS, CGYRO, TGLF, TGYRO, NEO, BoTorch, GPyTorch, ASTRA, TRANSP, JINTRAC, GX, TANGO, GENE, TRINITY, QuaLiKiz, PyTorch, GKW, NCLASS, EPED, TORIC, FPPMOD, MAESTRO, FreeGS, TEQ, RAPIDS, STEP, FUSE, OMFIT, NUBEAM, TORAX}
\newcommand{\MITIMfusion}{\texttt{MITIM-fusion}\xspace}

\newcommand{\maineq}{
\begin{aligned}
    \mathbf{z}^* = \argmax_{z_{m} \in [z^L_{m},z^U_{m}]} h
    &\left(\left\{
        \sum_{k\in k_\text{tr}}
            g^{\text{tr}}_{m,k}(\mathbf{z}) \cdot \widehat{F}^{\text{tr}}_{m,k}(h^{\text{tr},k}_{m}(\mathbf{z}))
    \right. \right.
    \\
    &\left. \left.
        - \sum_{k\in k_\text{tar}}
            h^{I}_m\!\left(g^{\text{tar}}_{m,k}(\mathbf{z}) \cdot \widehat{f}^{\text{tar}}_{m,k}(h^{\text{tar},k}_{m}(\mathbf{z}))\right)
    \right\}_{\forall m}\right)
\end{aligned}
}

\begin{document}

\title[MAESTRO workflow]{Accelerating integrated modeling with surrogate-based optimization: the \MAESTRO workflow}

\author*[1]{\fnm{P.} \sur{Rodriguez-Fernandez}}\email{pablorf@mit.edu}


\author[1]{\fnm{N. T.} \sur{Howard}}

\author[1]{\fnm{J.} \sur{Hall}}

\author[1]{\fnm{A.} \sur{Saltzman}}

\author[1,4]{\fnm{A.} \sur{Martin-Sanabria}}

\author[1]{\fnm{A.} \sur{Ho}}

\author[1]{\fnm{G.} \sur{Snoep}}

\author[1,4]{\fnm{J.} \sur{Pimentel-Aldaz}}

\author[2]{\fnm{C.} \sur{Holland}}

\author[1]{\fnm{M.} \sur{Muraca}}

\author[1,4]{\fnm{P.} \sur{de Lara Montoya}}

\author[1]{\fnm{K.} \sur{Yanna}}

\author[1]{\fnm{A. E.} \sur{White}}

\author[3]{\fnm{T.} \sur{Body}}

\author[3]{\fnm{A. J.} \sur{Creely}}

\author[3]{\fnm{J. C.} \sur{Hillesheim}}

\author[3]{\fnm{P. B.} \sur{Snyder}}

\affil[1]{\orgdiv{MIT Plasma Science and Fusion Center}, \city{Cambridge}, \state{MA}, \country{USA}}
\affil[2]{\orgdiv{University of California San Diego}, \city{La Jolla}, \state{CA}, \country{United States}}
\affil[3]{\orgdiv{Commonwealth Fusion Systems}, \city{Devens}, \state{MA}, \country{United States}}
\affil[4]{\orgdiv{Universidad Politecnica de Madrid}, \city{Madrid}, \country{Spain}}

\abstract{
This paper introduces the \MAESTRO workflow, that enables the coupling of the \PORTALS framework \cite{Rodriguez-Fernandez2024Nucl.Fusion_Enhancing} with external solvers for the plasma equilibrium, pedestal physics, divertor constraints and heating.
The surrogate-based optimization nature of the transport solver is ideally suited for external coupling, allowing efficient steady-state predictions of plasma profiles with full physics models.
Improvements in the surrogate modeling of quasilinear transport models with \PORTALS are presented, which enable the efficient handling of discontinuities in the transport fluxes that can arise from numerical issues or physical instabilities with extreme stiffness.
The combination of physics-informed methods and advanced numerical techniques allows the \MAESTRO workflow to provide accurate and efficient predictions of steady-state plasma profiles, which are critical for fusion reactor design and optimization.
}

\keywords{
PORTALS, surrogate optimization, uncertainty quantification, quasilinear, optimization, transport, turbulence, integrated modeling
\\
\\
\textit{Submitted for publication}
}

\maketitle

\section{Introduction}
\label{sec:Introduction}

Surrogate-based optimization has been demonstrated to be a powerful tool for the efficient prediction of steady-state, multi-channel flux-matched plasma profiles using $\delta$f  turbulent transport models with the \PORTALS framework \cite{Rodriguez-Fernandez2024Nucl.Fusion_Enhancing}.
The formulation of the multi-channel transport equations as a residual minimization problem \cite{Candy2009Phys.Plasmas_Tokamak} allows for the exploitation of surrogate-based and Bayesian optimization techniques.
These methods are particularly useful when dealing with expensive, black-box simulations, such as those performed with $\delta$f gyrokinetic codes that need to be coupled to the evolution of the background distribution function. 
They are also efficient to use with reduce transport models, as they have the potential to handle discontinuous and stiff behavior of the transport fluxes more efficiently than traditional finite differences based methods, as it will be presented in this work.

A critical component of self-consistent steady-state predictions of plasma profiles is the consideration of changes in the magneto-hydrodynamic (MHD) equilibrium and pedestal performance that occur as the plasma pressure evolves, as well as heating deposition and safety factor profiles that are consistent with the temperature and density of the background plasma.
Furthermore, in reactor-relevant regimes, the evolution of the pressure profile of energetic ions can significantly impact the MHD equilibrium and the background turbulence \cite{Siena2023Nucl.Fusion_Predictions}.
These changes must be accommodated during the numerical steps towards steady-state (e.g., in \ASTRA \cite{Pereverzev2002IPP-Rep.Max-Planck-Inst.FurPlasmaphys._ASTRA} or \TRANSP \cite{Breslau2018Comput.Softw.USDOEOff.Sci.SCFusionEnergySci.FES_TRANSP}) or as part of an external iteration loop (e.g. \STEP \cite{Lyons2023PhysicsofPlasmas_Flexible}, \FUSE \cite{Meneghini2024_FUSE}).

This work presents the first coupling of the \PORTALS framework with external modules that provide solutions to the MHD equilibrium (Grad-Shafranov solver), pedestal physics (peeling-ballooning stability), heating deposition (wave physics and Monte-Carlo fast ion codes), edge physics (divertor detachment models), fusion products (alpha slowing down and ash dynamics), and safety factor profile (current diffusion with a model for sawtooth instabilities) to provide low-computational cost, self-consistent solutions that are useful for fusion reactor design and optimization.
The coupling of these modules is referred to as the \textbf{M}odular and \textbf{A}ccelerated \textbf{E}ngine for \textbf{S}imulation of \textbf{T}ransport and \textbf{R}eactor \textbf{O}ptimization (\underline{\MAESTRO}) workflow, available in the open-source \MITIMfusion repository \cite{mitim}.

\subsection{Motivation}
In recent years, the plasma modeling community has developed several integrated modeling frameworks that couple multiple physics modules to provide self-consistent simulations of tokamak plasmas. The \MAESTRO workflow builds upon these efforts, but the integration of surrogate-based optimization techniques within \PORTALS presents a number of advantages that motivated its development:
\begin{itemize}
    \item The loose coupling of the different physics modules within both \PORTALS itself and the \MAESTRO workflow allows for modularity and flexibility in the choice of solvers for each physics component, requiring virtually no modification to the original codes. This enables the easy integration (in a black-box manner) of turbulent transport models of arbitrary fidelity, such as \TGLF \cite{Staebler2007Phys.Plasmas_theorybased}, \GX \cite{Mandell2022_GX} and \CGYRO \cite{Candy2016J.Comput.Phys._highaccuracy}.
    \item The use of \PORTALS, in contrast to time-independent Newton solvers and time-dependent partial differential equation solvers, is capable of finding steady-state solutions with a minimal number of evaluations of the transport fluxes, which is particularly advantageous when the addition of higher physics fidelity increases the computational cost of each evaluation.
    This is relevant when considering the need for nonlinear gyrokinetics (when quasilinear descriptions are not sufficient), when higher resolutions are required (e.g., number of basis functions in eigensolvers), or the inclusion of more fields (e.g., electromagnetic effects and multiple ion species).
    \item The steady-state formulation in \PORTALS directly targets the stationary solution of the transport equations. In time-dependent transport solvers, reaching steady state requires evolving the plasma profiles over many energy confinement times ---a requirement that becomes particularly burdensome for high confinement time scenarios.
    The strong nonlinearity of turbulent fluxes with respect to local gradients makes the coupled transport equations numerically stiff, forcing these solvers to use very small time steps relative to the confinement time and greatly multiplying the total number of flux evaluations required to reach stationarity.
    High-performance reactor regimes are particularly challenging in this context, with small $\rho^*=\rho_s/a$, where $\rho_s$ is the ion sound Larmor radius and $a$ is the minor radius.
    Particle (density) transport is also especially challenging in this context: the particle confinement time is typically several times longer than the energy confinement time, meaning density profiles take significantly longer to equilibrate in a time-dependent simulation. \PORTALS sidesteps all of these issues by treating the steady-state condition as a direct optimization target, when the goal is to find the stationary solution of the transport equations.
    \item The \MAESTRO workflow reduces the waste of computational resources by enabling the re-utilization of previously computed transport fluxes during the convergence process of each \PORTALS iteration \cite{Rodriguez-Fernandez2024Nucl.Fusion_Enhancing,Rodriguez-Fernandez2024PhysicsofPlasmas_Core}, particularly when the equilibrium and heating modules introduce small changes to the background profiles between iterations, or when the only feedback loop is through updates of the boundary conditions (such as in pedestal-core transport modeling loops).
    \item The use of independent radial grids for the transport solver portion (\PORTALS) of the \MAESTRO workflow reduces the needs for high performance computing resources when used with quasilinear models.
    Strong radial averaging of transport coefficients, often required in traditional integrated modeling frameworks \cite{Pereverzev2002IPP-Rep.Max-Planck-Inst.FurPlasmaphys._ASTRA,Breslau2018Comput.Softw.USDOEOff.Sci.SCFusionEnergySci.FES_TRANSP}, is no longer necessary, enabling the use of coarser grids.
    These coarser radial grids that can be handled by single compute nodes and laptop computers.
    This is also advantageous for the reduction of human-in-the-loop time during the setup of simulations, as it reduces the need for extensive convergence studies and checks (e.g., radial grid resolution, time step size, numerical diffusivity, flux matching).
\end{itemize}

A growing body of work accelerates plasma performance optimization by replacing expensive physics models with pre-trained neural-network surrogates. Such approaches are powerful when applicable, but efficient optimization with full physics models remains essential whenever validated surrogates are unavailable, the physics models themselves still require validation, or the offline cost of surrogate training cannot be justified ---conditions under which accurate predictions of plasma performance would otherwise remain out of reach.
The framework presented in this work is complementary to such efforts, and can be used to generate training data for surrogate models.

The rest of this paper is organized as follows.
Section \ref{sec:PORTALS} provides a brief overview of the \PORTALS transport solver, its formulation as a surrogate-based optimization problem, and the advantages of this approach when used with quasilinear turbulent transport models.
Section~\ref{sec:MAESTRO} describes the \MAESTRO workflow and its components, including the coupling strategy between \PORTALS and the external physics modules.
Section~\ref{sec:Application} presents an application of the \MAESTRO workflow to the prediction of steady-state plasma profiles in compact fusion power plant scenarios.
We must note that the efficiency of \PORTALS to handle multi-channel profile predictions with quasilinear models is the foundation of the integrated modeling workflow in \MAESTRO and therefore is described in detail in this paper, while the other components of the workflow are described at a higher level, as they are not the main focus of this work.

\section{The \PORTALS solver}
\label{sec:PORTALS}
The \PORTALS transport solver \cite{Rodriguez-Fernandez2022Nucl.Fusion_Nonlinear,Rodriguez-Fernandez2024Nucl.Fusion_Enhancing} is an extension of the \TGYRO code \cite{Candy2009Phys.Plasmas_Tokamak} that leverages Bayesian optimization techniques to find the steady-state solution of the multi-channel, multi-radial transport equations.
This paper expands upon previously published work on the \PORTALS framework by providing a more detailed description of the mathematical formulation of the surrogate-based optimization problem, as well as the techniques implemented to increase the robustness and efficiency of the solver.
As derived in Appendix~\ref{app:GeneralizedTransportSolver}, the steady-state solution of the transport equations can be found by maximizing a scalar function that quantifies the level of stationarity:
\begin{equation}
    \maineq
    \label{eq:PORTALSoptimization}
\end{equation}
where the set of $\mathcal{H} = \{h, h^I_m, h^{\text{tr}}_{m,k}, h^{\text{tar}}_{m,k}, g^{\text{tr}}_{m,k}, g^{\text{tar}}_{m,k}\}$ transformations are nonlinear functions that aid in the training and optimization process, and $\widehat{F}^{\text{tr}}_{m,k}$ and $\widehat{f}^{\text{tar}}_{m,k}$ are the normalized (in their native units) outputs of transport and target models, respectively.
$\mathbf{z}$ is the set of local gradients of the transport channels.
$m$ is the index that runs over the radial locations and channels, while $k$ runs over the different transport and target flux components.
The rest of the notation is explained in detail in Appendix~\ref{app:GeneralizedTransportSolver}.
In the context of this work, ``transport'' refers to the flux components that are associated with turbulent and neoclassical transport, while ``target'' refers to the flux components that are associated with sources and volumetric sinks.
Throughout this paper, a subscript $\forall j$ (or $\forall c$, $\forall m$, $\forall k$) denotes the collection over all values of that index, while a plain index denotes a fixed value; mixed subscripts such as $z_{j,\forall c}$ indicate a slice at fixed $j$ over all channels.

If the models of transport fluxes and target flux densities are written as a single, multi-output model $\mathbf{F}=\{\widehat{F}^{\text{tr}}_{m,k}, \widehat{f}^{\text{tar}}_{m,k}\}_{\forall m, \forall k}$, and the input space of normalized gradients is represented by $\mathbb{X} \subseteq \mathbb{R}^m$, then the notation can be simplified as:
\begin{equation}
\label{eq:eq:PORTALSsimple}
\begin{aligned}
    \mathbf{z}^* = \argmax_{\mathbf{z}\in\mathbb{X}}
    v
    \left(\mathbf{F}\right)
\end{aligned}
\end{equation}
where the scalar function $v$ includes all transformations required to map the individual flux calculations to the stationarity metric. 
Standard \PORTALS simulations typically involve the evaluation of $N_r\times N_c\times[k_{\text{tr}}+k_\text{tar}]$ fluxes, where $N_r$ is the number of radial locations, $N_c$ is the number of channels simulated, $k_{\text{tr}}$ the number of transport models (e.g. turbulence and neoclassical), and $k_\text{tar}$ is the number of sources and sinks (e.g. heating, energy exchange, alpha power and radiation).

As explained in Ref.~\cite{Rodriguez-Fernandez2024Nucl.Fusion_Enhancing}, the optimization problem is high dimensional, with $m$ input parameters ($N_r\times N_c$).
A key aspect of the \PORTALS solver is the use of surrogate models to approximate the transport fluxes and target flux densities.
This way, with the transformations $h^{\text{tr},k}_{m}$ and $h^{\text{tar},k}_{m}$, the entire dimensionality of the problem is reduced to the parameters that each flux component calculation requires (e.g. $2\times N_c$ for the transport fluxes when the transport models are not sensitive to second or higher order profile derivatives). 
We leverage Gaussian Processes (GPs) as surrogate models for the transport fluxes and target flux densities $\mathbf{F}$.

Because of the nonlinear transformations $\mathcal{H}$ introduced to simplify GP training, the resulting acquisition functions generally lack closed-form analytical expressions and must therefore be evaluated via Monte Carlo approximations.
This makes \PORTALS a direct application of composite Bayesian optimization \cite{Astudillo2019_Bayesian}, which applies when the underlying function is black-box, derivative-free, and expensive to evaluate, while the outer transformation is cheap.
In this setting, and contrary to standard Bayesian optimization techniques, we leverage the direct information retrieved from the transport fluxes and target flux densities evaluations ($\mathbf{F}$) and the direct knowledge of the nonlinear transformations $v$ to guide the optimization process more efficiently than directly modeling the objective function $v$ with a single Gaussian process (GP).

\subsection{On uncertainty quantification in deterministic quasilinear models}
\label{sec:convergence}

The GPs used in \PORTALS assume that each of the transport fluxes, $F^{tr}$, as a function of input parameters, $\textbf{x}$, follow independent joint Gaussian distribution functions:
\begin{align}
	F^{tr}(\textbf{x}) \sim \mathcal{G}\mathcal{P}(\mu(\textbf{x}),\mathcal{K}(\textbf{x},\textbf{x}'))
\end{align}
where $\mu$ is the mean function and $\mathcal{K}$ is the covariance kernel function that describes the correlation between different points.

It is also assumed that the transport flux evaluation at each point is corrupted by heteroscedastic Gaussian noise with zero mean and a known variance $\sigma_y^2$:
\begin{align}
    \label{eq:dev}
	y^{tr}(\textbf{x}) = F^{tr}(\textbf{x}) + \mathcal{N}(0,\sigma^{2}_{y}(\textbf{x}))
\end{align}

This formulation allows for the quantification of the uncertainty in the transport model predictions, which is particularly useful when dealing with noisy transport flux evaluations, such as those obtained from gyrokinetic initial value solvers.
The simulations performed with nonlinear \CGYRO \cite{Candy2016J.Comput.Phys._highaccuracy} using the \PORTALS framework have shown exceptional computational efficiency, often reaching steady-state solutions with as low as 10 (or less) profile evaluations (e.g. Ref.~\cite{howard_2026_arc}).
Such high efficiency is hypothesized to be a result of the ability of the GP surrogates to effectively handle the noise in the limited time averaging of transport flux evaluations, increasing robustness near regions of marginal stability. It is also aided by the dimensionality reduction techniques implemented, which fully leverage the local nature of the transport flux evaluations, while maintaining a global, integrated view of the transport equation in the surrogate optimization process.

Simulations with quasilinear transport models based on eigenvalue solvers, such as \TGLF \cite{Staebler2005Phys.Plasmas_GyroLandau,Staebler2007Phys.Plasmas_theorybased} and \QuaLiKiz \cite{Bourdelle2005PlasmaPhys.Control.Fusion_Turbulent}, present a different set of challenges for transport solvers.
While they do not rely on time averaging of turbulent fluctuations to obtain transport fluxes to communicate to the transport solver, they can exhibit discontinuous behavior as a function of the input parameters due to either numerical convergence issues of the eigenvalue solver or discontinuous transitions that result from the saturation rules employed in the quasilinear model.
These discontinuities can appear, for example, in transition boundaries where several linear modes compete for dominance, or when the saturation rules try to mimic physical nonlinear effects in a simplified manner (e.g., zonal flows or cross-scale coupling).

\begin{figure}[h!]
    \centering
    \includegraphics[width=0.9\textwidth]{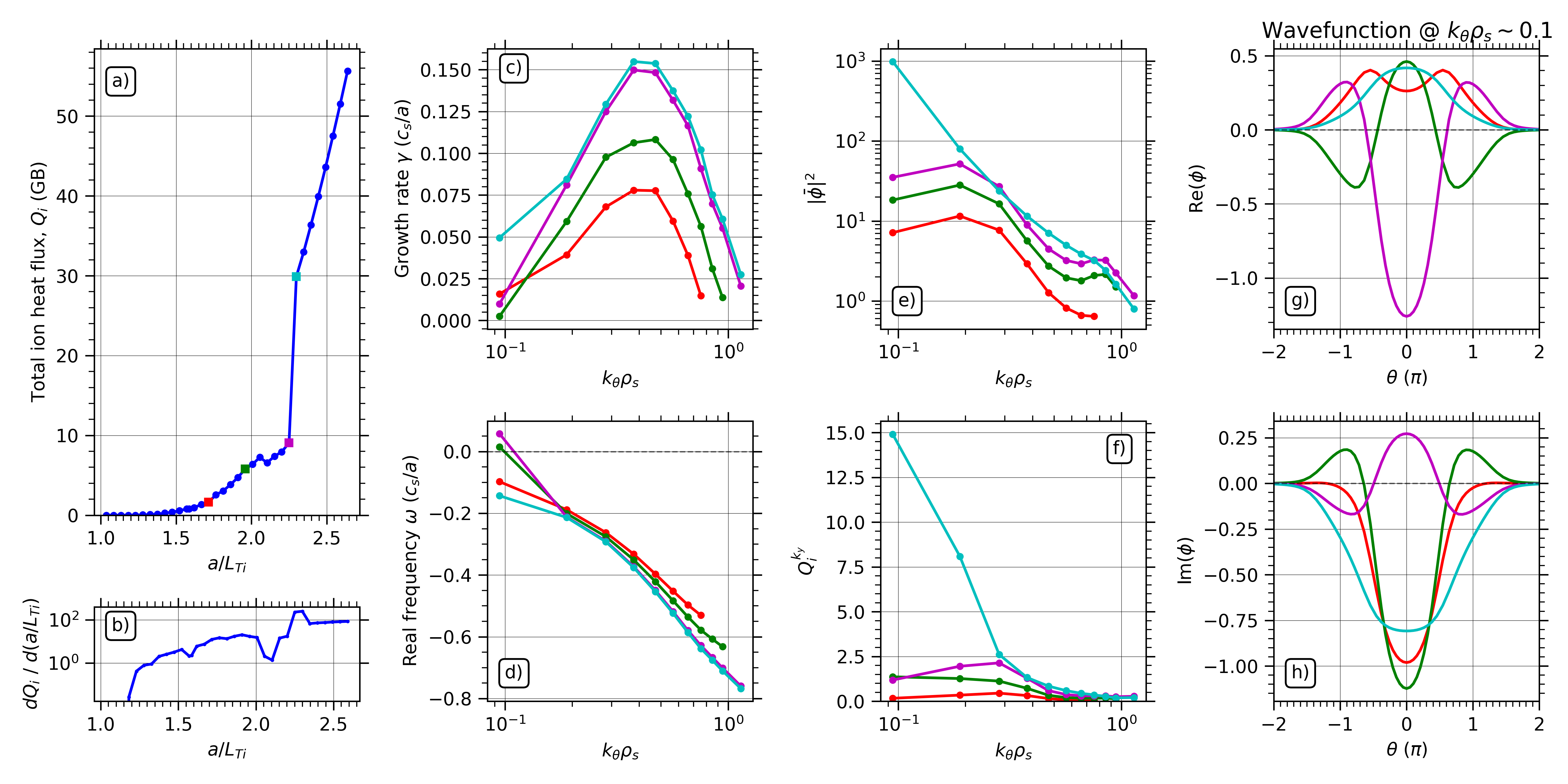}
    \caption{Example of discontinuity in \TGLF-SAT3 transport flux evaluations as a function of $a/L_{Ti}$.
    a) Ion heat flux as a function of $a/L_{Ti}$.
    b) Derivative of the ion heat flux with respect to $a/L_{Ti}$.
    c) Growth rate and d) real frequency spectra of the dominant linear mode for four different values of $a/L_{Ti}$ indicated by colors in panel a).
    e) Electrostatic potential intensity and f) ion heat flux spectra.
    g) Real and h) imaginary parts of the eigenmode associated with the dominant linear mode at $k_\theta\rho_s=0.1$.
    Each color refer to specific points in the $a/L_{Ti}$ scan, as indicated in panel a).
    }
    \label{fig:tglf_discontinuity}
\end{figure}

Figure~\ref{fig:tglf_discontinuity} shows an example of the discontinuous behavior that can be observed in transport evaluations with \TGLF-SAT3 \cite{Dudding2022Nucl.Fusion_new} as a function of the normalized ion temperature gradient, $a/L_{Ti}$, for fixed values of all other input parameters.    
This example corresponds to a SPARC Primary Reference Discharge \cite{Creely2020J.PlasmaPhys._Overview,Rodriguez-Fernandez2020J.PlasmaPhys._Predictions} plasma at $\rho_{tor}=0.85$ (square root of normalized toroidal flux), and shows a jump of $\sim300\%$ in the ion heat flux when the ion temperature gradient is increased by only $\sim2\%$, for one speicific value of $a/L_{Ti}$ (Figure~\ref{fig:tglf_discontinuity}a).
This behavior, which leads to order of magnitude changes in the local sensitivity of the transport flux to small changes in the input parameter (\ref{fig:tglf_discontinuity}b), is associated with a change in the dominant linear mode from electron to ion direction at the lowest wavenumber simulated of $k_\theta\rho_s=0.1$, where $\rho_s$ is the ion sound Larmor radius and $k_\theta$ is the poloidal wavenumber.
This results in a significant change in the mode structure (Figure~\ref{fig:tglf_discontinuity}g-h) that leads to a large increase in the electrostatic potential intensity at low wavenumbers (Figure~\ref{fig:tglf_discontinuity}e), which in turn results in a significant contribution to the ion heat flux (Figure~\ref{fig:tglf_discontinuity}f).

This type of discontinuities--- either due to numerical issues of the model, or physical instabilities with extreme stiffness--- can pose challenges for traditional transport solvers that rely on finite differences to compute the Jacobian of the transport fluxes with respect to the background profiles, as they can lead to inaccurate or unstable solutions. Solvers in recent years have relied on techniques such as radial averaging of the transport coefficients to smooth out these discontinuities, resulting in the need to run with a \textit{higher-than-needed} radial resolution.

The surrogate-based optimization techniques employed in \PORTALS are well-suited to handle these discontinuities, by reformulating the existence of numerical discontinuities as an additional source of uncertainty in the transport flux evaluations, as explored next.
We must note that this work does not intend to provide solutions to solve the discontinuous behavior of quasilinear models, but rather to provide techniques to mitigate their impact in the convergence behavior of transport solvers.

\begin{figure}[h!]
    \centering
    \includegraphics[width=0.9\textwidth]{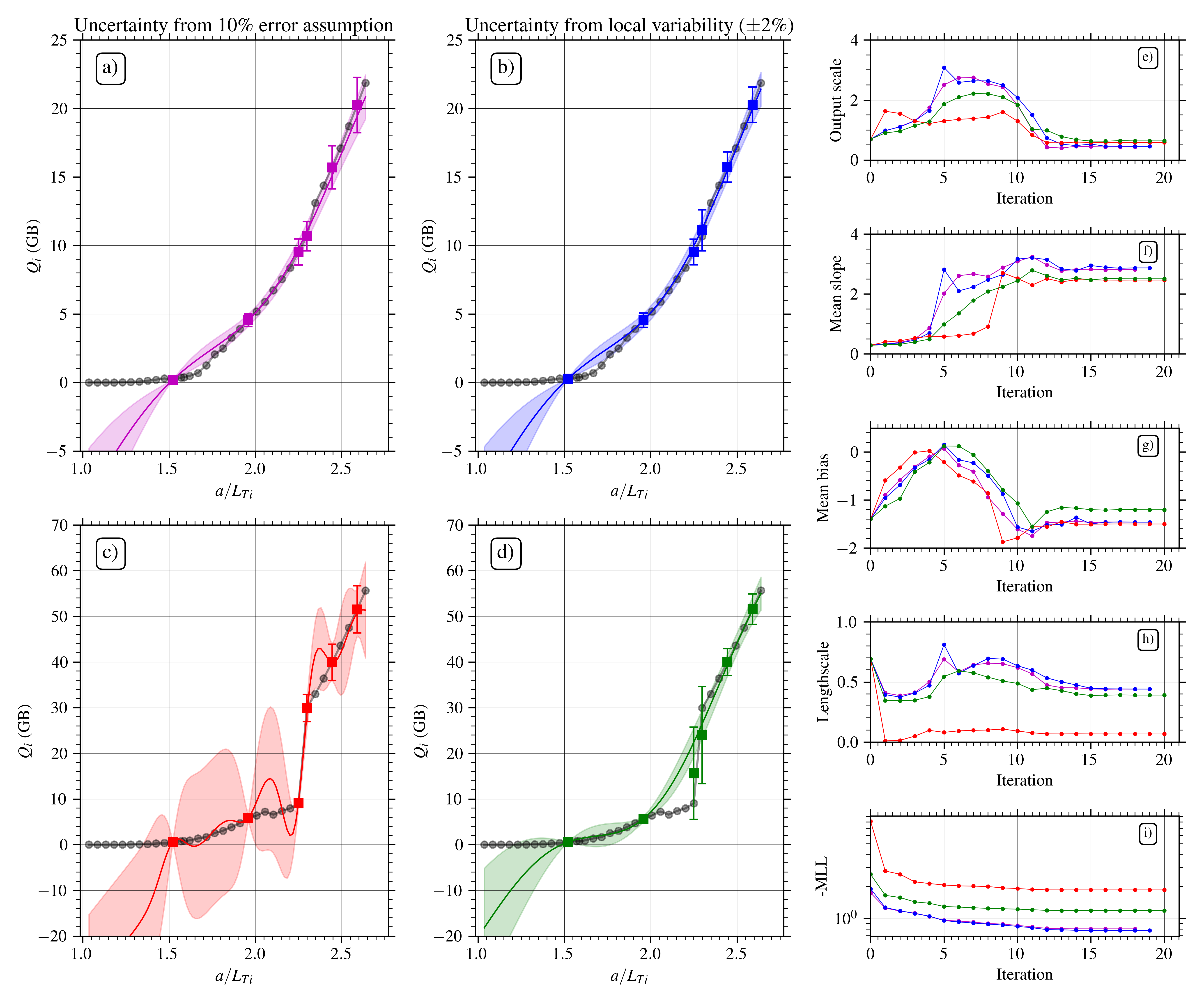}
    \caption{
        Behavior of the GP surrogate models (mean and one standard deviation) for well-behaved (a) and discontinuous (c) transport flux evaluations, when fitted assuming a fixed Gaussian noise of $10\%$ standard deviation in a subset (indicated in squares) of the \TGLF gradient scan.
        (b) and (d) show the same cases but when using the uncertainty quantification technique presented in Section~\ref{sec:convergence}, resulting in a better-behaved surrogate model.
        Evolution of hyperparameters during GP fitting (in normalized input space and standardized output space): (e) output scale, (f) slope of linear mean function, (g)  bias of linear mean function, and (h) RBF kernel length-scale.
        (i) Evolution of the marginal log-likelihood.
    }
    \label{fig:tglf_gp}
\end{figure}

Figure~\ref{fig:tglf_gp}a and c depict the behavior of a Gaussian Process surrogate model with a Radial Basis Function (RBF) kernel and linear mean function fitted with a fixed Gaussian noise of $10\%$ standard deviation for (a) a well-behaved and (c) a discontinuous  transport flux evaluation, respectively.
Figure~\ref{fig:tglf_gp}c corresponds to the same case as Figure~\ref{fig:tglf_discontinuity}. The uniform length-scale of the kernel is unable to properly capture the sharp transition in the transport flux evaluation while maintaining accuracy in the smooth regions of the transport curve, leading to large uncertainties in the surrogate model predictions and oscillatory behavior.
As shown in Figure~\ref{fig:tglf_gp}h, the length-scale associated with the discontinuous case (red) quickly takes very small values during the fitting process.
A surrogate model fitted with such a naive approach would be unable to provide accurate predictions of the transport fluxes, leading to poor convergence behavior in \PORTALS and potentially large number of transport flux evaluations to resolve the discontinuities.

In \PORTALS, the turbulent transport flux evaluations are treated as low-dimensional, local functions that are decoupled from the global transport equations.
Each local transport flux can be fully described by a set of input parameters that include the normalized logarithmic gradients (e.g., $a/L_T = -\frac{1}{a}\frac{dT}{dr}$) of the background profiles, $z_{j,c}$, as well as the local values of the profiles themselves, $y_{j,c}$, at each radial location, resulting in: $F^{tr}_{j,c}(z_{j,\forall c},y_{j,\forall c})$ \cite{Rodriguez-Fernandez2024Nucl.Fusion_Enhancing}, where $j$ and $c$ refer to the radial location and channel of the transport flux evaluation, respectively.
In the case of a standard three-channel ($T_e$, $T_i$, $n_e$) prediction, the maximum six-dimensional models for $Q_e$, $Q_i$ and $\Gamma_e$ can be written as: $F^{tr}_{j,c}( a/L_{Te}, a/L_{Ti}, a/L_{ne}, T_e, T_i, n_e )$.
This set of input parameters can be further transformed into a more physically intuitive set of dimensionless parameters that are commonly used in turbulent transport and that are direct drivers of the underlying microinstabilities, such as the normalized collisionality, $\widehat{\nu}_{ei}$, the temperature ratio, $T_i/T_e$ and the electron beta, $\beta_e$.
This new complete set retains the same dimensionality while improving the interpretability of the surrogate models, resulting in: $F^{tr}_{j,c}( a/L_{Te}, a/L_{Ti}, a/L_{ne}, \widehat{\nu}_{ei}, T_i/T_e, \beta_e )$.

\begin{figure}[h!]
    \centering
    \includegraphics[width=0.9\textwidth]{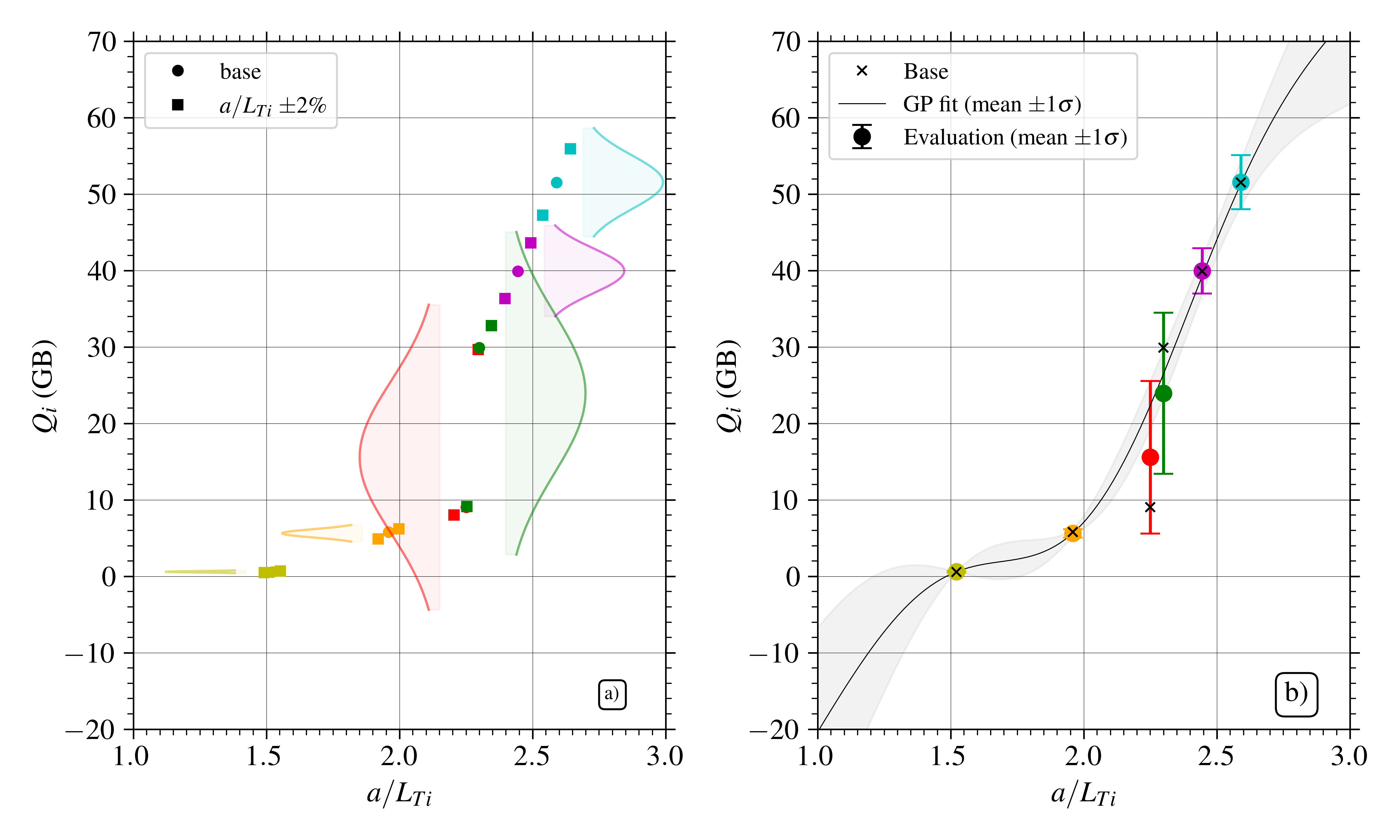}
    \caption{
        a) \TGLF evaluation of the ion heat flux as a function of $a/L_{Ti}$ (same as in Figure~\ref{fig:tglf_gp}c) for fixed values of all other input parameters. Each base evaluation is accompanied by additional evaluations at $\pm 2\%$ of $a/L_{Ti}$. Gaussian distributions are fitted to each set of evaluations to extract the mean and standard deviation of the transport flux evaluation at each input parameter point.
        b) Resulting GP surrogate model fitted to the mean transport flux evaluations with associated uncertainties from the standard deviations.
        Each color indicates different evaluations of the gradient scan.
    }
    \label{fig:tglf_scan}
\end{figure}

In order to quantify the uncertainty in each of \TGLF transport flux evaluations, we make the choice of scanning $\pm 2 \%$ around each of the input parameters used in a given transport flux evaluation. This results in a total of $(1+2\times 2\times N_c)$ transport flux evaluations per point, where $N_c$ is the number of channels being predicted (e.g., $N_c=3$ for $T_e$, $T_i$ and $n_e$).
As illustrated in Figure~\ref{fig:tglf_scan} for the one-dimensional case, this sampling strategy allows us to capture the local variability of the transport fluxes around each input parameter point, which can then be used to estimate the variance in the transport flux evaluation.
From the distribution of evaluations, we compute the mean and standard deviation of the transport fluxes, and that is reported to \PORTALS as the transport flux evaluation and its associated uncertainty, respectively.

The application of this uncertainty quantification technique results in significant improvements in the surrogate model predictions, as shown in Figure~\ref{fig:tglf_gp}d as compared to Figure~\ref{fig:tglf_gp}c for the discontinuous transport flux evaluation case.
It is worth mentioning that the GP posterior mean does not interpolate the training points exactly because each observation carries finite uncertainty (a non-zero noise/variance term in the likelihood). The GP balances fidelity to the data against the prior smoothness, so points with larger uncertainties pull the mean less strongly.
We acknowledge that this technique increases the computational cost of each transport flux evaluation, as multiple \TGLF simulations need to be performed at each input parameter point. This is also similar to radial averaging approaches, but with the advantage that every other parameter is held constant and that the information extracted from the transport flux evaluations is fully utilized in the surrogate model and subsequent integrated modeling steps with \MAESTRO.

The technique illustrated in this section can only capture limited sources of uncertainty in the transport flux evaluations, where the effect of each input parameter is assumed to be independent from the others\footnote{One could run as many combinations of inputs and as many relative variations from the mean as needed, but the cost of the uncertainty quantification technique would quickly grow to unmanageable levels.}.
To alleviate to some degree this limitation, we have implemented an additional check that, if a previously evaluated sample from the uncertainty quantification set is found in the \PORTALS database to be within the same relative distance (e.g., $2\%$ in the high-dimensional ball around the sampling point), its transport flux evaluation is used as part of the uncertainty quantification set of the new point.

\subsection{Stopping criteria for \PORTALS}
\label{sec:stopping_criteria}

The use of uncertainty quantification--- either the one presented in Section~\ref{sec:convergence} for eigenvalue solvers or the one in Ref.~\cite{Rodriguez-Fernandez2024Nucl.Fusion_Enhancing} for initial value solvers--- allows for the implementation of more robust stopping criteria for the \PORTALS framework that has advantages over traditional scalarized-residual-based, or iterations-based criteria.
Having access to uncertainty estimates in the transport flux evaluations, we can use them to determine when the surrogate models have converged to a solution that is within the uncertainty bounds of the transport flux evaluations.
To this end, we aim to find a scalar metric that captures the overall goodness-of-fit in the transport flux evaluations across all channels and radial locations.

A metric for the overall level of agreement between simulation and experiment was proposed in Ref.~\cite{Ricci2011Phys.Plasmas_Methodology}, and referred in subsequent works as the \textit{Ricci validation metric} (e.g., Ref.~\cite{MolinaCabrera2023PhysicsofPlasmas_Isotope}).
This metric is an average across measurements of the level of agreement, weighted with the quality of each observable, with 0 representative of perfect agreement and 1 of perfect disagreement.

\begin{figure}[h!]
    \centering
    \includegraphics[width=1.0\textwidth]{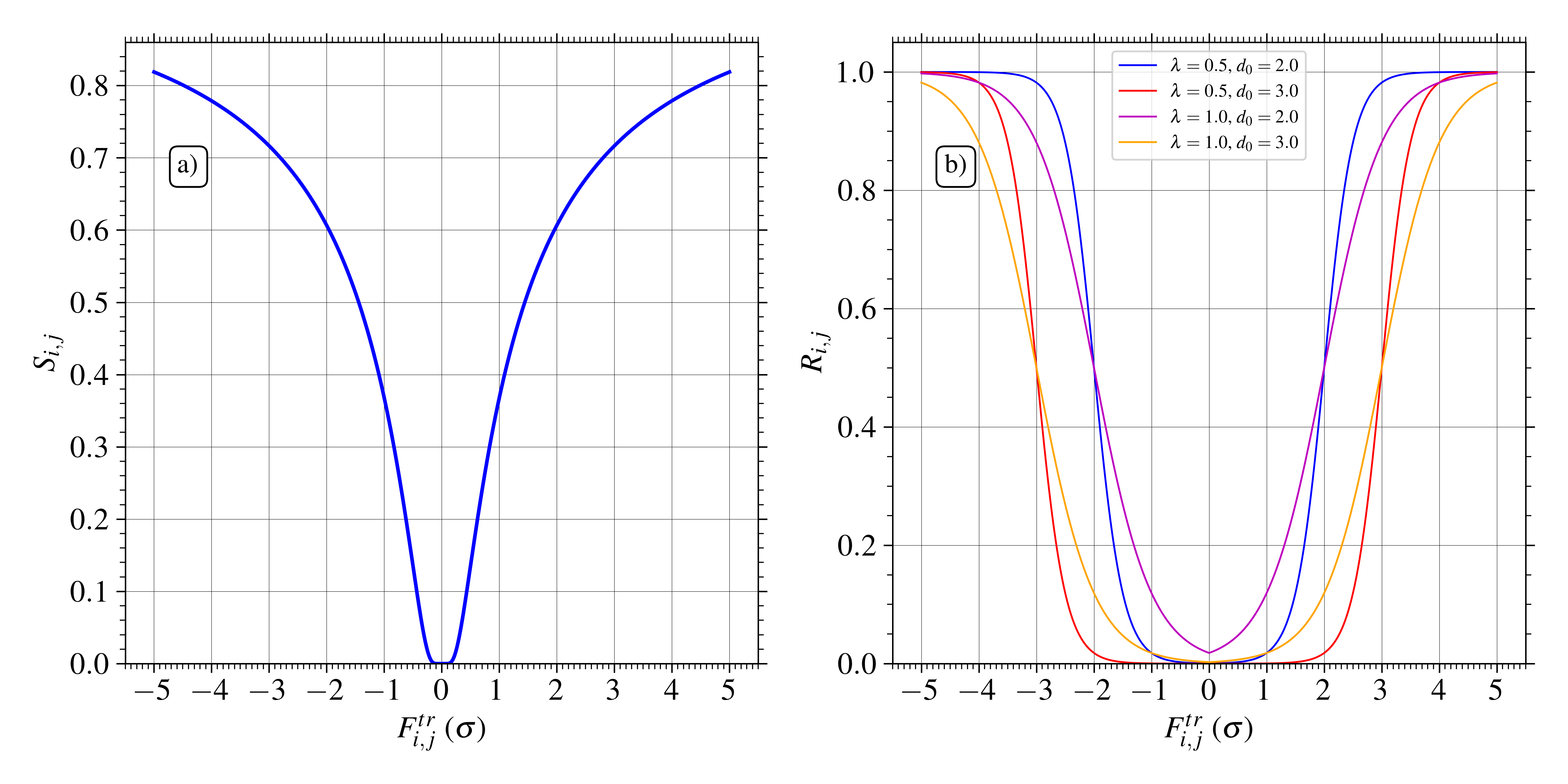}
    \caption{
        Ricci metric components for an example case with zero target flux with no uncertainty ($F^{\text{target}}_{j,c}=\Delta F^{\text{target}}_{j,c}=0.0$), and transport flux evaluation with unit uncertainty ($\Delta F^{\text{tr}}_{j,c}=1.0$).
        a) Quality factor $S_{j,c}$ and b) agreement factor $R_{j,c}$, as a function of transport flux evaluation, $F^{\text{tr}}_{j,c}$.
        The agreement factor is evaluated with different choices of the parameters $d_0$ and $\lambda$.
    }
    \label{fig:ricci}
\end{figure}

In this work, we adapt the Ricci validation metric to define a stopping criterion for \PORTALS based on the level of agreement between the target fluxes, $F^{\text{target}}_{j,c}$, and the transport fluxes, $F^{\text{tr}}_{j,c}$.
We consider all radii and channels to be weighted equally (hence with a constant hierarchy level), resulting in a metric of the form:
\begin{align}
    \chi_R = \frac{\sum_{j,c}R_{j,c}S_{j,c}}{\sum_{j,c}S_{j,c}}
    \label{eq:Ricci}
\end{align}
where $S_{j,c}$ quantifies the quality of the radius-channel residual evaluation:
\begin{align}
    S_{j,c} = \exp\left(
        -\frac{
            \Delta F^{\text{tr}}_{j,c}+\Delta F^{\text{target}}_{j,c}
            }{
            \left\|F^{\text{tr}}_{j,c}\right\|+\left\|F^{\text{target}}_{j,c}\right\|
            }
    \right)
\end{align}
where $\Delta F^{\text{tr}}_{j,c}$ and $\Delta F^{\text{target}}_{j,c}$ are the uncertainties of the transport and target flux evaluation respectively.
Note that the significance of $S_{j,c}$ resides in that more accurate evaluations will be weighted more ($S_{j,c}\rightarrow 1$) than inaccurate ones, preventing a situation where a very large uncertainty evaluation dominates the convergence metric because transport and target falls within their large uncertainties.
Figure~\ref{fig:ricci}a shows an example that illustrates that for values of the transport flux evaluation of the order of its uncertainty, $F^{\text{tr}}_{j,c}=1.0$ (relative uncertainty of $100\%$), the weight is $S_{j,c}\sim0.37$, as compared to $S_{j,c}\sim0.61$ when the transport flux evaluation is as high as $2$ times its uncertainty (relative uncertainty of $50\%$).

On the other hand, the normalized agreement is determined in the following form:
\begin{align}
    d_{j,c} = \sqrt{        
        \frac{            
                \left(
                    F^{\text{tr}}_{j,c} - 
                    F^{\text{target}}_{j,c}
                \right)^2
            }{
                \left(\Delta F^{\text{tr}}_{j,c}\right)^2 + 
                \left(\Delta F^{\text{target}}_{j,c}\right)^2
            }
        }
\end{align}
which enters in Equation~\ref{eq:Ricci} via the term $R_{j,c}$ that quantifies the level of agreement between transport and target:
\begin{align}
    R_{j,c} = \frac{1}{2}\left(
        1 + \tanh\left(
            \frac{d_{j,c}-d_0}{\lambda}
        \right)
    \right)
\end{align}
where $d_0$ and $\lambda$ are ad-hoc parameters that define the sharpness of the metric.
Figure~\ref{fig:ricci}b illustrates how $R_{j,c}$ approaches 0 when the transport approaches target ($F^{\text{target}}_{j,c}=0$ assumed for simplicity) within the uncertainty bounds, and approaches 1 when the transport flux evaluation is significantly different from target.
$\lambda$ provides a mechanism to tune the sharpness of the transition, while $d_0$ sets the offset.
In this work we have found that values of $d_0=2.0$ and $\lambda=0.5$ provide good behavior in practice, as a value of the agreement factor $R_{j,c}$ approaching 0.0 ($R_{j,c}\leq 0.02$ in the example of Figure~\ref{fig:ricci}) is obtained when the transport flux evaluation is within the combined uncertainty of transport and target ($F^{\text{tr}}_{j,c}\leq 1.0$), and an intermediate value of $R_{j,c}\sim0.5$ when the transport flux evaluation is within two times the uncertainty.

\begin{figure}[h!]
    \centering
    \includegraphics[width=1.0\textwidth]{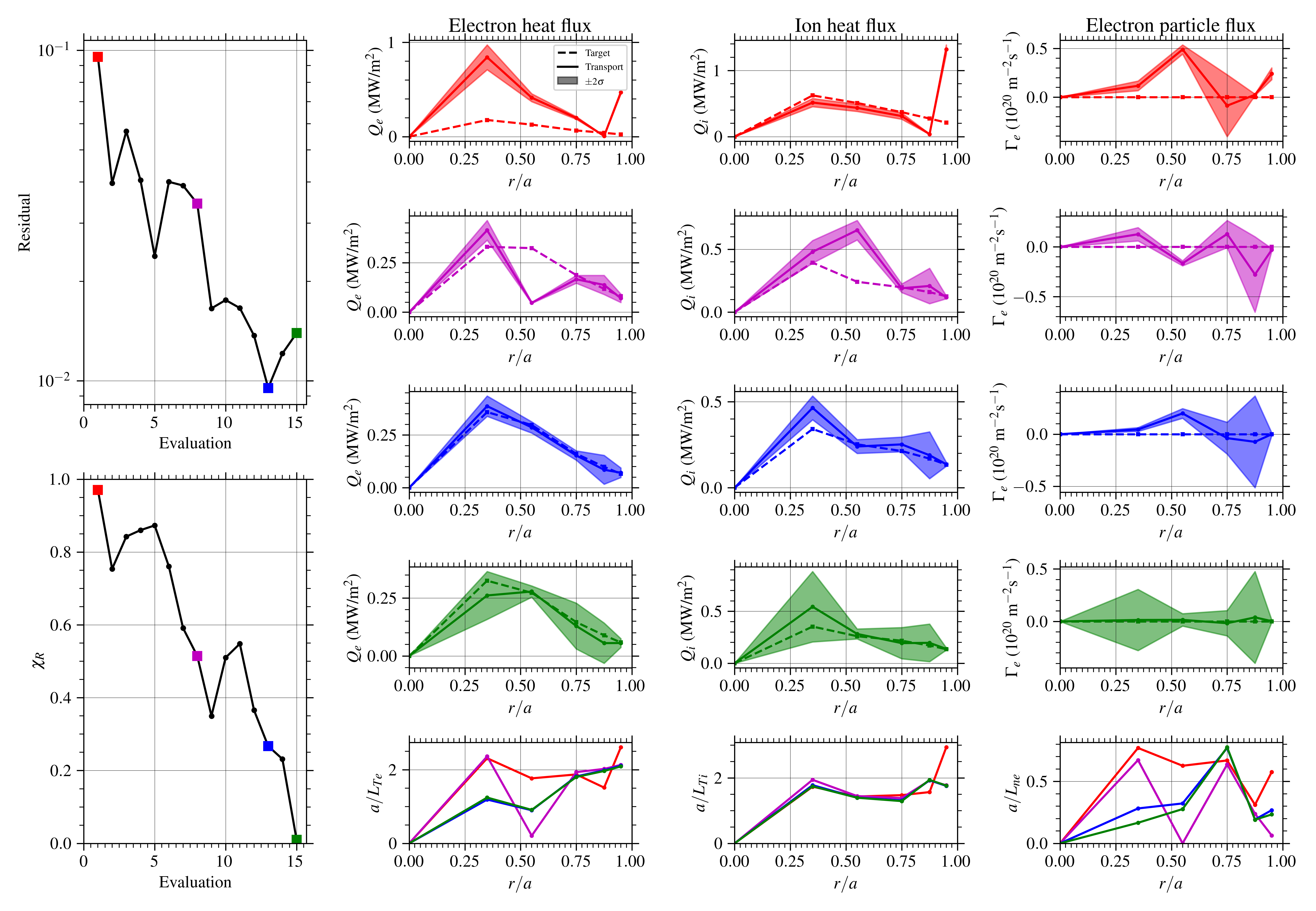}
    \caption{
        Evolution of the (upper left) $L_2$ residual and (lower left) Ricci metric during \PORTALS iterations.
        (rest) Flux-matching quality in electron heat flux, ion heat flux, and electron particle flux, and normalized logarithmic gradients at four selected iterations. Mean and 2 standard deviations of the transport flux evaluations are shown, and the target fluxes are shown as dashed lines.
        Colors (red, magenta, blue, green) indicate specific iteration numbers for better visualization.
    }
    \label{fig:portals_ricci}
\end{figure}

The Ricci metric, $\chi_R$, thus provides a global metric that can be used to characterize the flux-matching quality level.
Figure~\ref{fig:portals_ricci} shows an example of the evolution of the Ricci metric during \PORTALS iterations for a SPARC Primary Reference Discharge \cite{Creely2020J.PlasmaPhys._Overview,Rodriguez-Fernandez2022Nucl.Fusion_Overview} scenario using \TGLF-SAT3 \cite{Dudding2022Nucl.Fusion_new} as turbulent transport model, with alpha heating, energy exchange and radiation losses included self-consistently.
The $L_2$ residual (defined from the differences between transport and target fluxes at all channels and radii) and the the Ricci metric both evolve in a similar manner, decreasing as the iterations progress. However, as the plasma approaches flux-matching, the Ricci metric provides a more informative picture of the flux-matching quality, as it accounts for the uncertainties in the transport flux evaluations.
As shown when comparing the last three iterations the residual may increase even though the Ricci metric improves and is very close to zero. The associated flux-matching plots show that the fluxes are well within the uncertainty bounds of the transport flux evaluations for the last iteration, even if higher residuals are observed (e.g. a flux discontinuity or high stiffness of transport model has been reached).

This metric does provide the benefit of being more physically interpretable than a simple residual, and it allows to stop the \PORTALS iterations without requiring ad-hoc thresholds on the residual, number of iterations, or the magnitude of the updates in the normalized logarithmic gradients, which would otherwise need to be tuned for each specific case and many more iterations could be required to ensure convergence.
In the example shown in Figure~\ref{fig:portals_ricci}, a stopping criterion based on the residual alone would have required tuning for this specific case, and very likely would have required more iterations to ensure convergence than with the Ricci metric.
Furthermore, it removes the challenge of having to choose appropriate weights or transformations (such as transforming particle fluxes to convective energy fluxes to avoid disparate scales) for each channel and radius in the residual calculation, as the Ricci metric naturally weights each residual based on its absolute uncertainty.
In this specific example, 15 iterations were sufficient to attain convergence, demonstrating the efficiency of \PORTALS even when using quasilinear transport models with discontinuous behavior.

\subsection{Physics-guided improvements to the \PORTALS framework}
\label{sec:other_considerations}

Since its initial publication \cite{Rodriguez-Fernandez2022Nucl.Fusion_Nonlinear}, and subsequent improvements \cite{Rodriguez-Fernandez2024Nucl.Fusion_Enhancing,Rodriguez-Fernandez2024PhysicsofPlasmas_Core}, the \PORTALS framework has incorporated several features that improve its robustness and efficiency.

\subsubsection{Positive diffusion constraint}
\label{sec:positive_diffusion}

To avoid unphysical solutions during the convergence process, particularly during the early iterations when the surrogate models are still being built, \PORTALS has implemented the ability to impose constraints on the surrogate models.
An important constraint that has been implemented is the enforcement of positive diagonal transport coefficients in the construction of the GP mean functions.
For example, $\frac{\partial \mu_{Qi}}{\partial a/L_{Ti}}>0$, where $\mu_{Qi}$ is the mean function of the ion heat flux surrogate model, and $a/L_{Ti}$ is the normalized logarithmic gradient of the ion temperature.
This constraint is physically grounded in the fundamental principle that transport fluxes must increase with their conjugate thermodynamic driving forces, a connection to irreversible thermodynamics.

\begin{figure}[h!]
    \centering
    \includegraphics[width=0.9\textwidth]{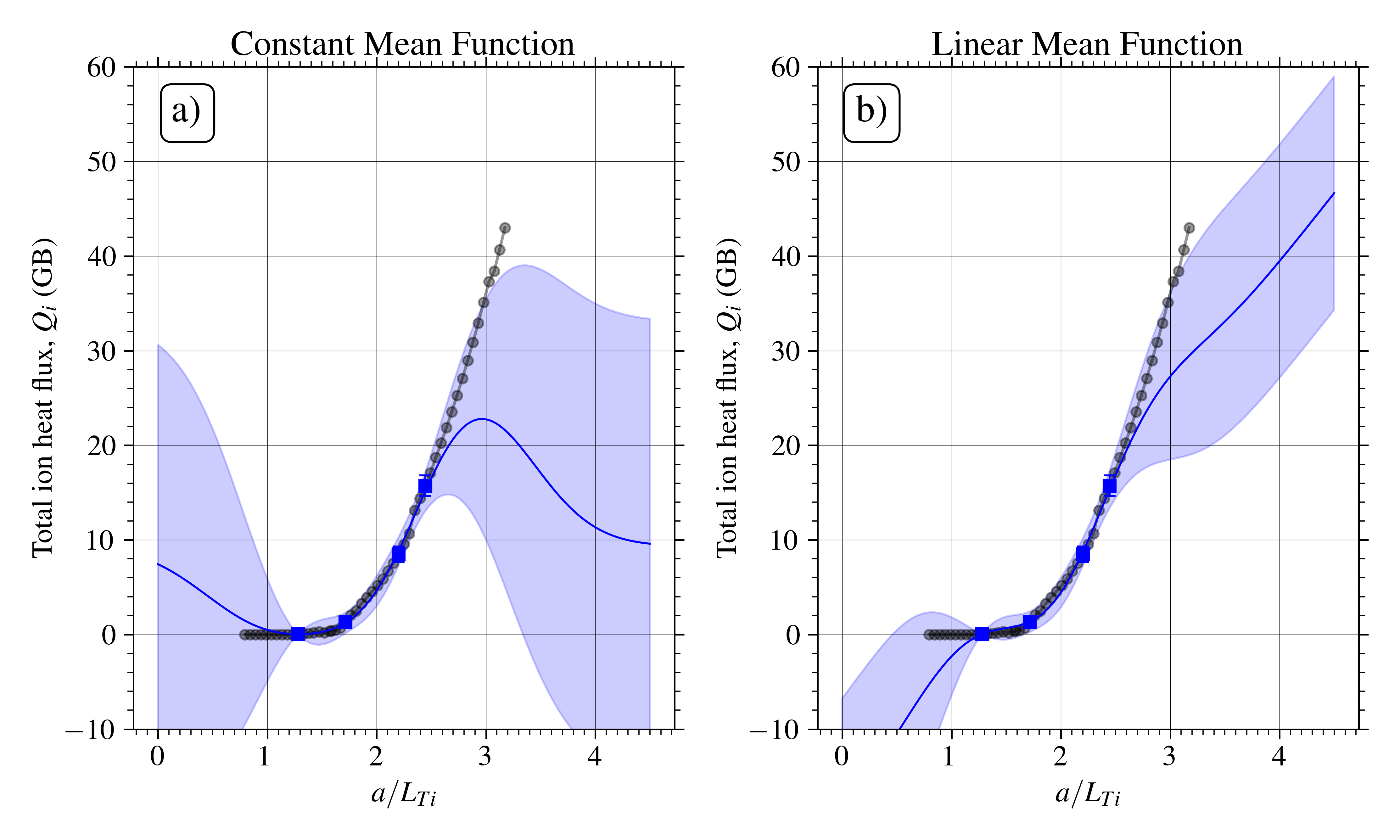}
    \caption{
        Example of ion heat flux as a function of $a/L_{Ti}$ for fixed values of all other input parameters in \TGLF.
        Effect of mean function choice in Gaussian Process surrogate models is shown when only four training points are used: (a) constant mean function, (b) linear mean function with positive diffusion constraint.
    }
    \label{fig:tglf_mean}
\end{figure}

This ``diffusive-like transport'' constraint has been shown to lead to significant improvements in the convergence behavior.
Standard GPs in Bayesian optimization workflows typically make the simplification of using constant (often zero) mean functions, letting the kernel handle the fit to data. However, in the context of transport solvers, this can lead to unphysical solutions in regions with lower or higher gradients than those previously evaluated, as shown in Figure~\ref{fig:tglf_mean}. The incorporation of domain-knowledge-based mean functions, such as linear or quadratic functions that enforce positive transport coefficients, helps to mitigate this issue.
We note that this constraint does not preclude the existence of transport flux evaluations that do not strictly follow diffusive behavior (e.g., transport dominated by sheared flows, or convective pinches in particle transport), as such behavior would be captured by the kernel of the GP.

\subsubsection{Dynamic Simple Relaxation (DSR)}
\label{sec:DSR}

As shown in Appendix~\ref{sec:Composite}, the evaluation of the mean and standard deviation of the combined GP ($N_r\times N_c\times N_m=5\times 3 \times 3 = 45$ invididual surrogates) with 15 training points, on 1000 random points, has an average computational cost in the tens of miliseconds.
While this is relatively inexpensive, the requirement to perform many evaluations during the acquisition optimization workflow, including Jacobian calculations, or when Monte-Carlo methods are used to account for nonlinear objectives in the acquisition function \cite{Wilson2018Adv.NeuralInf.Process.Syst._Maximizing,Balandat2020_BoTorch}, can carry a non-negligible overall cost.

Leveraging the use of the (analytical) mean of the posterior distribution as the acquisition function\footnote{We note that this makes the acquisition optimization be equivalent to solving for the steady-state, flux-matching condition for the mean prediction of the surrogate models.}, we implemented yet another domain-knowledge based technique to accelerate the convergence behavior of \PORTALS: the use of a simple relaxation technique to update the normalized logarithmic gradients towards flux-matching.
In Ref.~\cite{Rodriguez-Fernandez2024Nucl.Fusion_Enhancing}, the simple relaxation technique was introduced, similar to the one used in the \TGYRO code \cite{Candy2009Phys.Plasmas_Tokamak}. The gradients were updated following:
\begin{align}
\label{eq:SR}
    z_{j,c}^{(i+1)} = z_{j,c}^{(i)}+\eta_{j,c}\frac{F^{\text{target}}_{j,c}-F^{\text{tr}}_{j,c}}{\sqrt{\left(F^{\text{target}}_{j,c}\right)^2+\left(F^{\text{tr}}_{j,c}\right)^2}}\cdot \lvert z_{j,c}^{(i)}\rvert
\end{align}
where $\eta_{j,c}$ determines the relative step in normalized logarithmic gradients, $z_{j,c}$.
Simple relaxation works particularly well with the linear mean functions described in Section~\ref{sec:positive_diffusion}, as the surrogate models are already biased towards diffusive transport behavior, and the relaxation steps help guide the optimization process towards flux-matching solutions.

This technique was found to be useful in improving the initialization strategy in \PORTALS (vs random training) and as a first-cut towards flux-matching in \TGYRO. However, the need to choose appropriate relaxation parameters, $\eta_{j,c}$, made it difficult to generalize its use, and exact convergence was often not achieved and required further optimization steps.

Here, we implement a modification to the simple relaxation technique, referred to as Dynamic Simple Relaxation (DSR), where the relaxation parameters are dynamically adjusted based on the convergence behavior of the transport equations.
Starting from a uniform initial value of $\eta_{j,c}=\eta_0$, the relaxation parameters are decreased by a factor of $\eta_{D}$ whenever the corresponding residual experiences oscillatory behavior, unless they are already below a minimum threshold $\eta_{min}$.

Oscillatory behavior is detected by monitoring the last $n_{osc}$ iterations of each residual, and checking if either of three conditions are met:
\begin{enumerate}
    \item The evolution is completely flat, i.e. no change in $n_{osc}$ iterations, checked by asessing if the standard deviation of the residual evolution is zero. This convergence criterion suggests that the system has reached a steady state, or the flux-matching solution is outside of the optimization bounds.
    \item There is a dominant frequency ($>30\%$ of the total power spectral density) in the Fourier transform of the residual evolution after removing the zero-frequency component.
    \item The residual evolution is dominated by high-frequency components (i.e., more than $50\%$ of the total power spectral density is contained in frequencies in the higher third of the frequency range).
\end{enumerate}

It is found that using the ad-hoc values $\eta_0=0.1$, $\eta_D=5$, $\eta_{min}=10^{-6}$ and $n_{osc}=100$ provides robust convergence behavior across a wide range of scenarios with no human tuning.
We note that this is a derivative-free method that works well with smooth surrogates.
It assumes the existence of a zero residual solution for each channel, a condition guaranteed by the positive diffusion constraint described in Section~\ref{sec:positive_diffusion}, except for cases where target fluxes increase more strongly than transport fluxes with increasing gradients (i.e., thermal instability).

\begin{figure}[h!]
    \centering
    \includegraphics[width=1.0\textwidth]{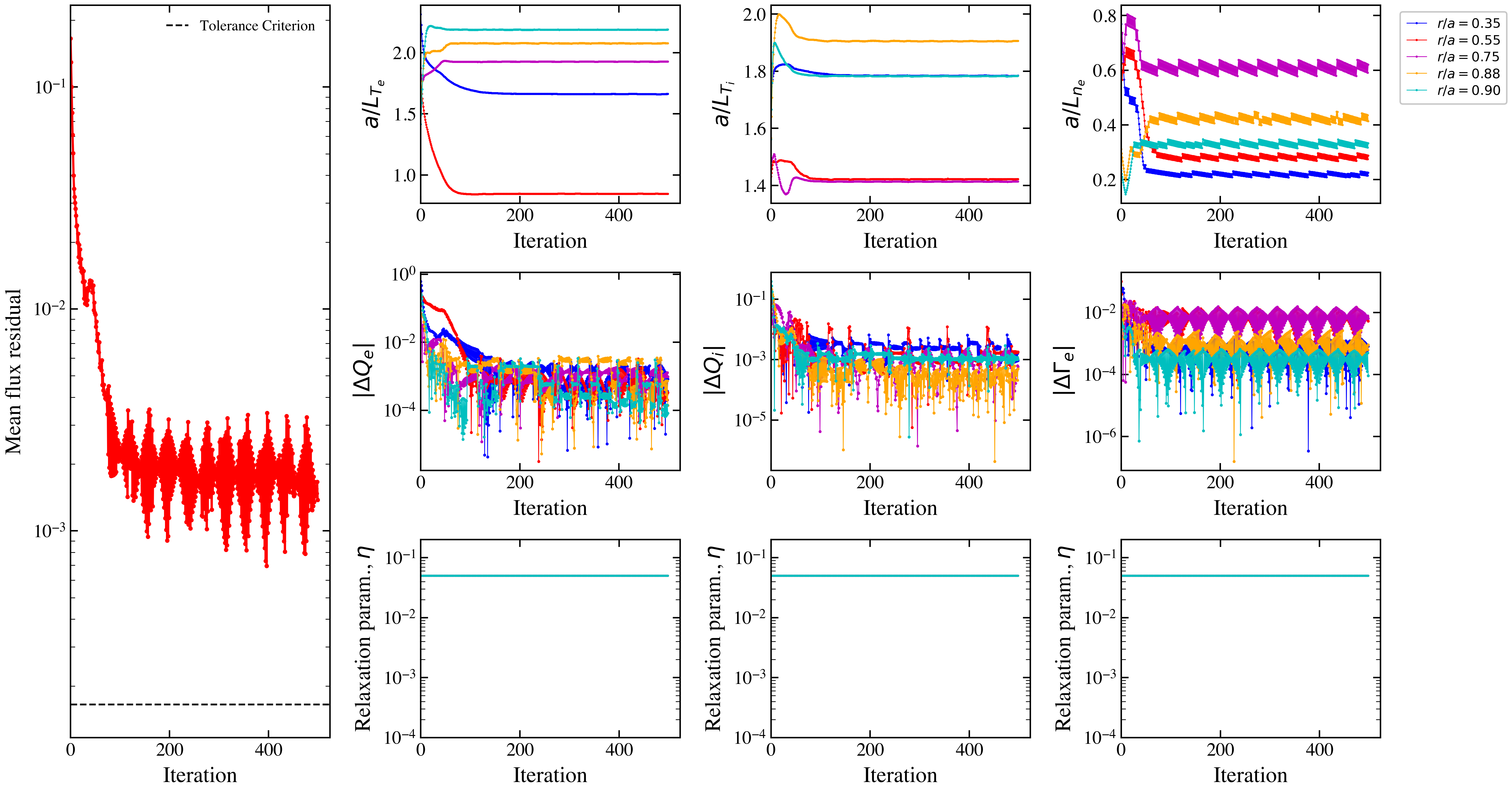}
    \caption{
        Evolution of the (left) mean flux residual, (top) gradient scale lenths, (middle) individual channel residuals, and (bottom) relaxation parameters during simple relaxation with fixed relaxation parameter of $\eta=0.05$. Colors indicate the different radii, with legend shown in the top right panel.
    }
    \label{fig:sr}
\end{figure}

\begin{figure}[h!]
    \centering
    \includegraphics[width=1.0\textwidth]{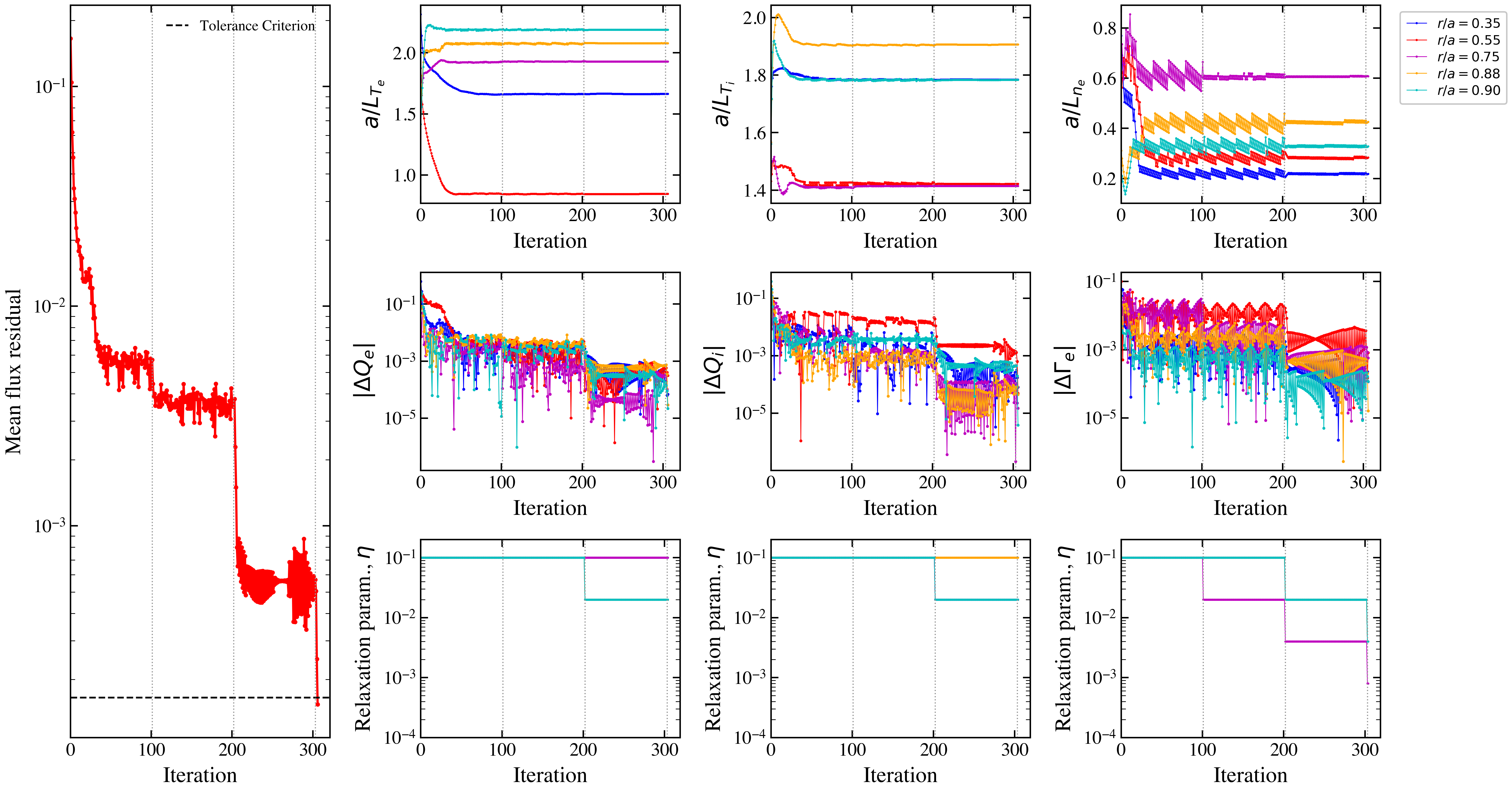}
    \caption{
        Evolution of the (left) mean flux residual, (top) gradient scale lenths, (middle) individual channel residuals, and (bottom) relaxation parameters using the dynamic simple relaxation technique explained in Section~\ref{sec:DSR}. Colors indicate the different radii, with legend shown in the top right panel. Dashed vertical lines indicate iterations where the relaxation parameters are reduced.
    }
    \label{fig:dsr}
\end{figure}

As an example of the benefits of DSR, Figure~\ref{fig:sr} shows the convergence behavior of the acquisition optimization in a \PORTALS simulation with a fixed relaxation parameter of $\eta=0.05$ for all channels and radii, while Figure~\ref{fig:dsr} shows the convergence behavior of \PORTALS with DSR.
The use of dynamic relaxation parameters allows for a more efficient convergence behavior, as the relaxation parameters are automatically reduced when oscillatory behavior is detected, preventing the system from stagnating and allowing it to converge to the flux-matching solution.
We note that situations may arise where, without adjustment of the relaxation parameters, the traditional simple relaxation technique may not be able to achieve convergence because it cannot resolve very stiff flux-gradient relationships.

The speed improvements described in Appendix~\ref{sec:Composite}, together with DSR, result in a flux-matching solution for the surrogates in $<$5 wall-clock seconds in a laptop for the example shown here.

\subsection{Application to flux-matching with TGLF}
\label{sec:benchmarks}

The fast, robust, and no-human in the loop convergence behavior of \PORTALS allows for the study of quasilinear model predictions in large databases of plasma scenarios without resorting to offline-trained surrogate models (e.g. \TGLF-NN) or low-fidelity physics models.
Here, we demonstrate this by constructing a database of 2048 generic tokamak fusion power plant (FPP) scenarios with varying plasma parameters as shown in Table~\ref{tab:database}, and performing flux-matching with \TGLF-SAT3 \cite{Dudding2022Nucl.Fusion_new} (including perpendicular magnetic fluctuations) as turbulent transport model and \NEO \cite{Belli2008PlasmaPhys.Control.Fusion_Kinetic} as neoclassical transport model at 6 radial locations: $r/a = 0.35, 0.55, 0.75, 0.875, 0.9, 0.94$.
In this exercise, only temperature profiles are evolved, and the input (subtracted radiation) power is externally imposed, with no alpha heating.
Magnetic equilibrium and safety factor profile are simply scaled up and down from a reference SPARC scenario to match desired geometry and kink safety factor, $q^{\star}$, values.

\begin{table}[h!]
    \centering
    \caption{Parameter values used to generate the generic tokamak FPP scenario database. Parameters such as the plasma current, $I_p$, are not directly specified, but are determined by the kink safety factor $q^{\star}$. Similarly, the density boundary condition comes from the choice of the fraction of the Greenwald density, $f_{G,top}$.
    Here, $top$ refers to the location of the boundary condition (e.g., $r/a=0.94$) and $sep$ the separatrix.
    }
    \label{tab:database}
    \begin{tabular}{ll}
        \toprule
        Parameter & Values \\
        \midrule
        $P_{in, tr}$ [MW] & $\{75,\ 150\}$ \\
        $R$ [m] & $\{3.5,\ 4.5\}$ \\
        $a$ [m] & $\{1.0,\ 1.2\}$ \\
        $B_T$ [T] & $\{8.0,\ 12.0\}$ \\
        $q^{\star}$ [-] & $\{3.0,\ 5.0\}$ \\
        $\kappa_{sep}$ [-] & $\{1.9,\ 2.1\}$ \\
        $\delta_{sep}$ [-] & $\{0.5,\ 0.7\}$ \\
        $f_{G,top}$ [-] & $\{0.7,\ 1.0\}$ \\
        Core $a/L_n$ [-] & $\{0.2,\ 0.6\}$ \\
        $T_{top}$ [keV] & $\{1.0,\ 3.0,\ 5.0,\ 8.0\}$ \\
        \bottomrule
    \end{tabular}
\end{table}

\begin{figure}[h!]
    \centering
    \includegraphics[width=1.0\textwidth]{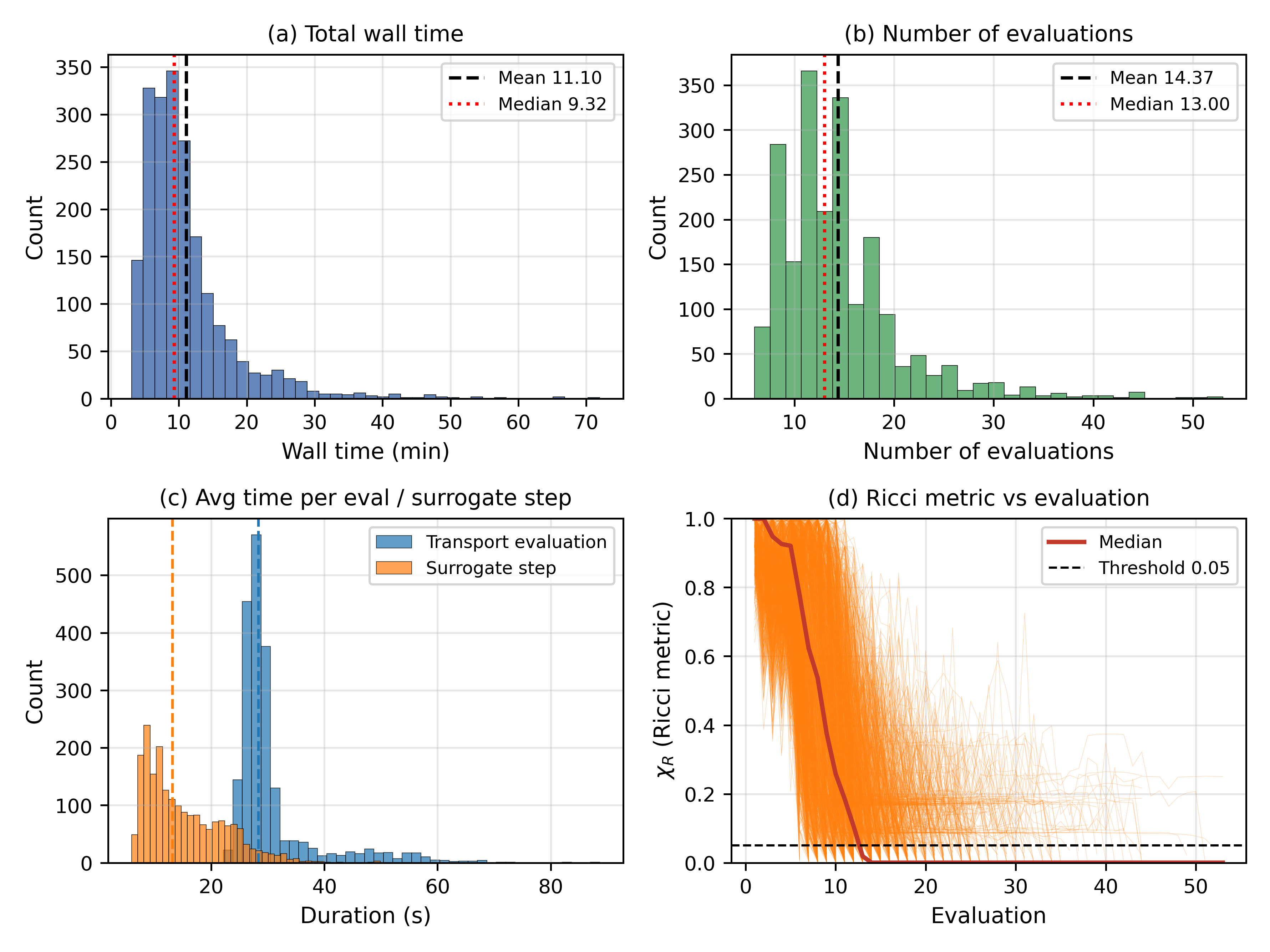}
    \caption{Statistics of the \PORTALS convergence behavior for the 2048 generic tokamak scenarios with \TGLF-SAT3 and \NEO, for the prediction of core temperature profiles.
    Each \PORTALS simulation is performed using 16 cores (on AMD EPYC cluster nodes) and with a maximum of 4 threads for surrogate operations to avoid over-threading for small operations (see Appendix~\ref{sec:Composite}).
    Histogram of the (a) wall-time and (b) number of evaluations required to achieve convergence for each scenario, with median values of 9 minutes and 13 evaluations, respectively.
    (c) Histogram of the the average time per evaluation consumed for the transport model evaluations and the surrogate model training and optimization, with median values of 28 seconds and 13 seconds, respectively.
    (d) Evolution of the Ricci metric per iteration for all cases.
    }
    \label{fig:database}
\end{figure}

Figure~\ref{fig:database} shows the statistics of the convergence behavior of \PORTALS. Converged solutions are obtained with only 13 transport evaluations on average.
Only 1.5\% of the cases failed to converge with the default settings of \PORTALS, despite the very wide range of plasma parameters explored.
These cases require further analysis to understand the reasons for the failure, but the low failure rate is a testament to the robustness of the \PORTALS framework, particularly when using quasilinear transport models with discontinuous behavior.
Time per evaluation is dominated by the transport model evaluations, which run instances of \TGLF and \NEO on 6 radial locations, and calculate the uncertainty of the \TGLF evaluation as described in Section~\ref{sec:convergence}, with a median value of 28 seconds per evaluation. This suggests that wall-time can be further reduced by allocating more than 16 cores per \PORTALS simulation, until the surrogate model operations (training and optimization) start to become the bottleneck (explored in Section~\ref{sec:Application}).

\section{The \MAESTRO workflow}
\label{sec:MAESTRO}

\MAESTRO is a loose-coupling integrated modeling workflow inspired by \OMFIT-\STEP \cite{Lyons2023PhysicsofPlasmas_Flexible}, where different physics modules are iterated until the plasma state does not change anymore.
In this paradigm, each physics module is called in sequence and treated as a black box: it receives the current plasma state as input and returns updated plasma quantities as output, without requiring internal knowledge of the other modules.
Self-consistency across physics domains---transport, equilibrium, heating, pedestal, and edge---is then achieved iteratively, with the outer loop repeating until the plasma state converges to a fixed point, i.e., changes between successive iterations fall below a prescribed tolerance.
This modular design allows the workflow to be readily extended with new physics modules without modifying existing ones, and enables flexible mixing and matching of solvers depending on the simulation requirements.

At the time of writing this manuscript, the main components of the \MAESTRO workflow are:
\begin{itemize}
    \item \underline{Heating, internal equilibrium, fusion products and current diffusion}: The \TRANSP \cite{Breslau2018Comput.Softw.USDOEOff.Sci.SCFusionEnergySci.FES_TRANSP} code ---with its singularity container version---, in its interpretive mode, is used as the main framework for modeling the internal equilibrium, wave heating (with \TORIC \cite{Brambilla1999PlasmaPhys.Control.Fusion_Numerical} and realistic antenna geometry), fast ion physics (with \NUBEAM \cite{Pankin2004Comput.Phys.Commun._tokamak}), current diffusion, and sawtooth modeling.
    \item \underline{Turbulent and neoclassical transport}: The \PORTALS framework is used as the time-independent transport solver, with the ability to use different transport models, such as \TGLF \cite{Staebler2007Phys.Plasmas_theorybased}, \GX \cite{Mandell2022_GX} and \CGYRO \cite{Candy2016J.Comput.Phys._highaccuracy} for turbulent transport, and \NEO \cite{Belli2008PlasmaPhys.Control.Fusion_Kinetic} for neoclassical transport. Analytical descriptions for the radiation, alpha heating and energy exchange channels are also included in \PORTALS (as described in Ref.~\cite{Rodriguez-Fernandez2022Nucl.Fusion_Nonlinear}), allowing for a self-consistent treatment of these channels in the transport solver.
    \item \underline{Edge modeling}: The \EPED \cite{Snyder2009Phys.Plasmas_Development} code is available as the pedestal model, to calculate pedestal pressure and width consistent with peeling-ballooning constraints, i.e. for ELMy H-mode plasmas.
    \MAESTRO is able to handle both the full version of \EPED, as well as neural network surrogate models trained on \EPED outputs.
    To calculate separatrix conditions, the Extended Lengyel model \cite{Body2025Nucl.Fusion_simple} can be used for the calculation of the separatrix temperature and the puffed impurity concentration required for detachment conditions.
    \item \underline{Separatrix modeling}: \MAESTRO can be initialized by providing the coordinates of the plasma separatrix or an equilibrium file ---obtained from experimental reconstructions or from free-boundary equilibrium solvers--- or it can be initialized using the \FreeGS \cite{freegs} free-boundary equilibrium solver, matching desired plasma quantities: major radius $R$, minor radius $a$, plasma current $I_p$, elongation $\kappa$ and triangularity $\delta$.
\end{itemize}

\begin{figure}[h!]
    \centering
    \includegraphics[width=1.0\textwidth]{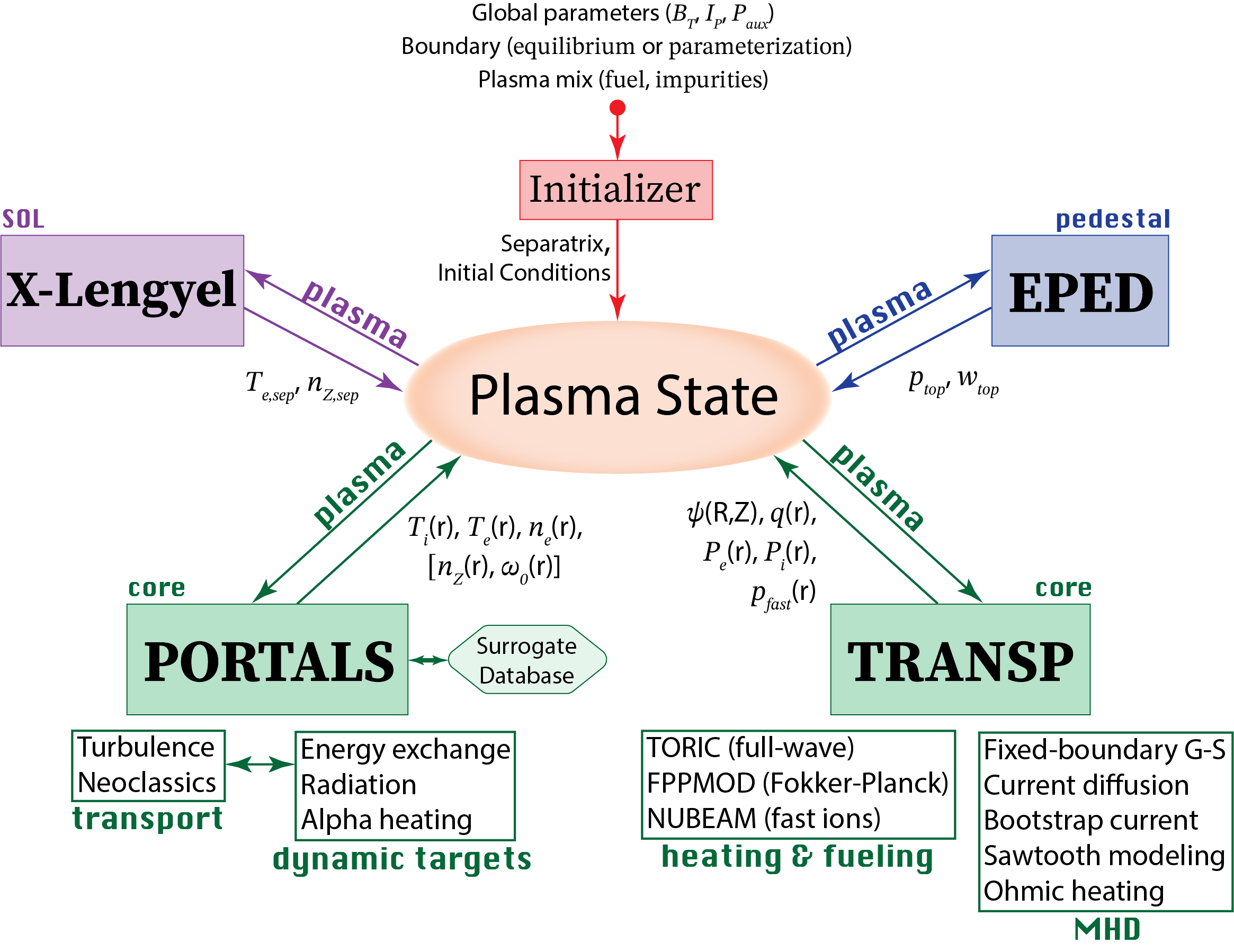}
    \caption{Schematic diagram of the \MAESTRO workflow.}
    \label{fig:MAESTRO_diagram}
\end{figure}

Figure~\ref{fig:MAESTRO_diagram} depicts a visual and simplified diagram of \MAESTRO \footnote{
    Allowing for a musical analogy, we view a \MAESTRO simulation as an orchestration process. Each outer-loop step is a ``\underline{beat}'': at every beat, transport, equilibrium, heating, pedestal, and edge modules contribute their own lines to the score. These lines are then re-voiced and balanced across successive beats, reducing mismatches much like resolving harmonic tension. Convergence is reached when the full ensemble settles into a stable cadence, with no section requiring further retuning from one beat to the next.
}.
Each module, interchangeable as per the needs for the specific simulation, modifies the plasma state depending on the physics modeled. 
This means that the current set of models can be readily expanded or switched out with ease, allowing for the inclusion of new physics models as they become available, and for testing the impact of different physics models on the predicted plasma performance.

The primary advantage of \MAESTRO is its speed to achieved steady-state solutions with \underline{full} physics models rather than machine learning surrogates.
We make here the distinction between ``full" and ``surrogate": surrogate models have recently become a driver of novel integrated modeling frameworks (such as \FUSE \cite{Lyons2023PhysicsofPlasmas_Flexible} and \TORAX \cite{Citrin2024_TORAX})), and although powerful, they rely on significant previous work training the high-dimensional space and, once trained, they have limited capabilities for testing numerical parameters and physics choices in the models.
The focus on delivering efficient methods with full physics models is motivated by the need to provide accurate predictions of plasma performance in regimes where surrogate models are not yet available, for which testing and validation of the physics models is still needed, and for which the computational cost of the offline training of surrogate models is not justified.

The time independent nature of the solver also means that it is agnostic to the transport time-scales and therefore simulating a reactor-like, burning plasma with long energy and particle confinement times, is not necessarily more computationally expensive than simulating a present-day device.
This makes \MAESTRO particularly advantageous when compared to solvers that need to bring the plasma to stationary conditions via solving a parabolic partial differential equation until the time evolution of stored energy, $dW/dt$, is small enough. Even when using smart time-stepping schemes, the often stiff nature of transport in burning regimes results in stringent computational requirements even when quasilinear models are used.

\section{Application to the prediction of compact fusion power plants}
\label{sec:Application}

In the following, we apply and describe the \MAESTRO workflow used to predict the plasma performance of FPP devices.
\MAESTRO became the tool of choice for the latest physics basis of the ARC device \cite{hillesheim_2026_arc_physics_basis,howard_2026_arc}, given its flexibility, robustness and speed. And it was verified against the \ASTRA \cite{Pereverzev2002IPP-Rep.Max-Planck-Inst.FurPlasmaphys._ASTRA} transport solver for a subset of cases.

Simulations are initialized using a full equilibrium reconstruction obtained using \FreeGS \cite{freegs}. The initial guess of kinetic profiles is calculated using pedestal + constant normalized gradient scale lengths in the core, following Ref.~\cite{Saltzman2025Nucl.Fusion_Impact}. Pedestal top pressure is calculated using \EPED \cite{Snyder2009Phys.Plasmas_Development}, and the core gradients ($a/L_{T}=a/L_{T_e}=a/L_{T_i}$ and $a/L_{n}$) are such that the plasma matches the initial guess of a global normalized pressure $\beta_N=2.0$ and density peaking of $\nu_{ne}=1.3$.
The separatrix from the \FreeGS reconstruction, together with the guess of kinetic profiles, is input to a \TRANSP \cite{Pankin2004Comput.Phys.Commun._tokamak} simulation to calculate the internal equilibrium and current diffusion.
The internal equilibrium is calculated with the \TEQ \cite{LoDestro1994Phys.Plasmas_GradShafranov} Grad-Shafranov code in fixed-boundary mode.
The Porcelli sawtooth model \cite{Porcelli1996PlasmaPhys.Control.Fusion_Model} is used to prevent the safety factor from dropping below 1 in the core, with parameters as done in Ref.~\cite{Rodriguez-Fernandez2020J.PlasmaPhys._Predictions}, and the Hager model \cite{Hager2016Phys.Plasmas_Gyrokinetic} is used for the calculation of the bootstrap current.
For this initial step, no wave heating nor fast ion physics is included, so that simulations can run for long times without significant computational cost.
The goal of this first step is to provide reasonable quasi-steady-state current density profiles and safety factor profile, so that the core transport simulations are not affected by the otherwise arbitrary choice of safety factor and magnetic shear profiles in \FreeGS.

Once the initial \TRANSP simulation finishes, a more complete \TRANSP simulation is performed, including the same physics as before but with the addition of wave heating with \TORIC \cite{Brambilla1999PlasmaPhys.Control.Fusion_Numerical} and \FPPMOD \cite{Hammett1986_Fast} to evolve the minorities distribution self-consistently, and fast ion physics with \NUBEAM \cite{Pankin2004Comput.Phys.Commun._tokamak}.
This second step is used to provide the heat deposition to the bulk plasma and the fast ion distribution.
The use of \NUBEAM also makes \MAESTRO readily applicable to simulations of plasmas heated with neutral beam injection.

\EPED is then used to calculate the pedestal pressure and width, consistent with the newly calculated plasma state (updated $\beta_N$ with fast ion pressure), and \PORTALS is used to find the flux-matching solution for the core plasma, with \TGLF-SAT3 as turbulent transport model and \NEO for neoclassical transport.
Electron temperature ($T_e$), ion temperature ($T_i$) and electron density ($n_e$) profiles are evolved at $r/a = 0.35, 0.45, 0.55, 0.65, 0.75, 0.875, (1-w_{top})$, where $w_{top}$ is the width of the pedestal top predicted by \EPED.
It is found that for this first instance of \PORTALS, running with fixed sources (radiation and alpha heating from \TRANSP) and for a few iterations (15 iterations) is sufficient to find a starting condition to iterate further with another \TRANSP step.

As a final stage, a set of 3 pairs of \EPED-\PORTALS steps are performed, which in this case are sufficient for the pedestal and core plasma to converge to the same solution. The pedestal pressure from \EPED affects the core plasma through the boundary conditions, and the core plasma affects the pedestal through the $\beta_N$ input to \EPED.
Because the equilibrium, currents and species concentrations are fixed, the \EPED-\PORTALS iterations can benefit from the reutilization of the surrogate models \cite{Rodriguez-Fernandez2024PhysicsofPlasmas_Core}, which significantly reduces the number of transport evaluations required to find the flux-matching solution for the core plasma, and therefore the overall computational cost of the \MAESTRO workflow.

\begin{figure}[h!]
    \centering
    \includegraphics[width=1.0\textwidth]{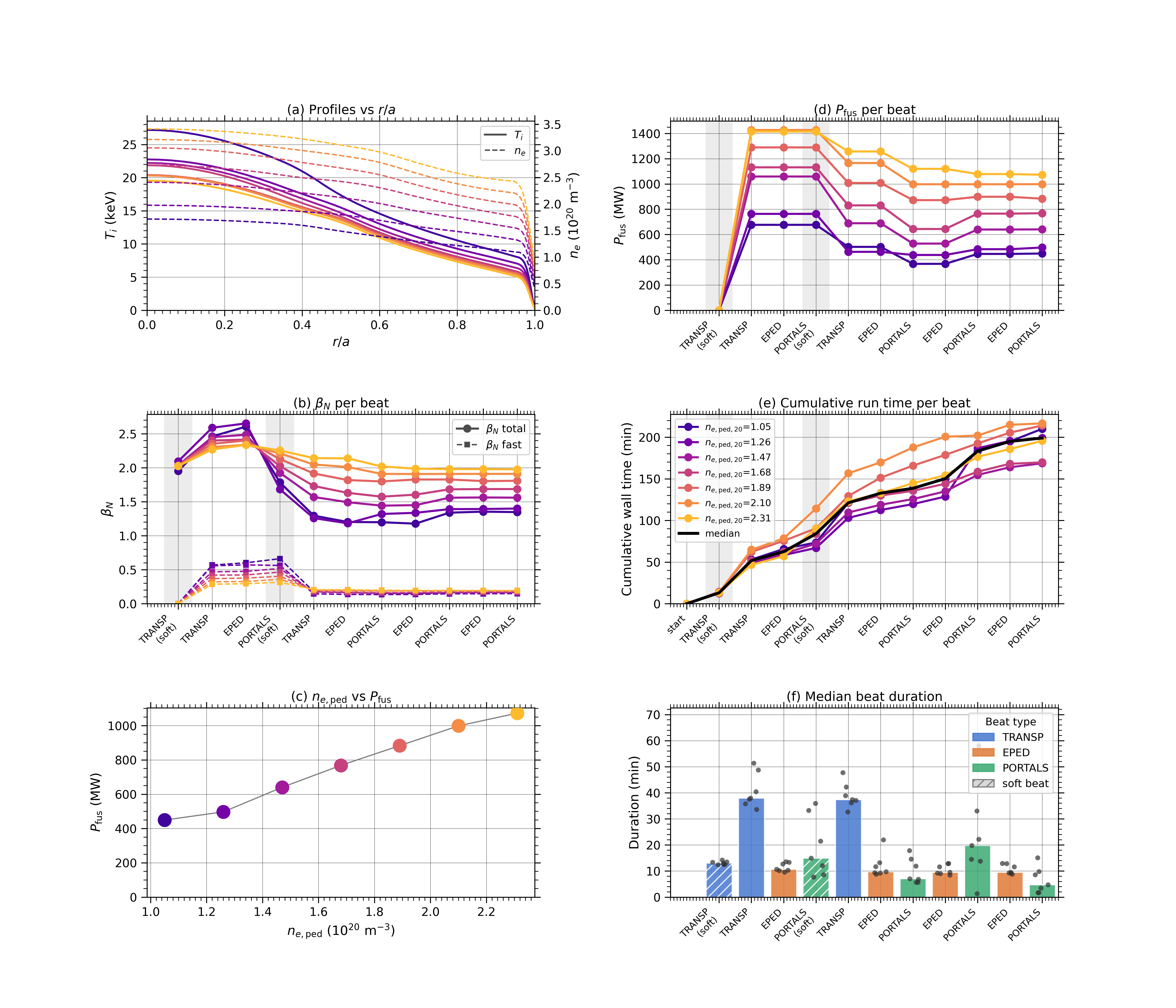}
    \caption{Scan of pedestal density, $n_{e,\text{ped}}$, for ARC V3A \cite{hillesheim_2026_arc_physics_basis, howard_2026_arc}, performed on AMD EPYC cluster nodes with 32 cores.
    (a) Steady-state ion temperature and electron density profiles.
    (b) Evolution of $\beta_N$ and its fast ion component during the \MAESTRO workflow.
    (c) Steady-state fusion power as a function of the pedestal density.
    (d) Evolution of fusion power and (e) cumulative run time during the \MAESTRO workflow.
    (f) Median beat duration, with scatter points indicating the duration for each case.
    }
    \label{fig:maestro_application1}
\end{figure}

Figure~\ref{fig:maestro_application1} shows the results and timing analysis of the \MAESTRO workflow for a scan of pedestal density, $n_{e,\text{ped}}$ (at constant $n_{e,\text{sep}}/n_{e,\text{ped}}$), for ARC V3A \cite{hillesheim_2026_arc_physics_basis, howard_2026_arc}.
In the \MAESTRO terminology, each iteration of the workflow (i.e., each call to the physics modules) is referred to as a ``beat", where ``soft" beats refer to the first \TRANSP step with no wave heating nor fast ion physics and the first \PORTALS step with fixed sources and limited iterations.
In this work, pedestal density is an input to the framework, and, as expected, it has a significant impact on the predicted fusion power (\ref{fig:maestro_application1}c), increasing overall performance (\ref{fig:maestro_application1}b), with nearly unchanged core profile shapes (\ref{fig:maestro_application1}a).
Simulations in \MAESTRO are stopped when sensitive quantities such as the fusion power and $\beta_N$ do not change from beat to beat (\ref{fig:maestro_application1}d), which happens after a median time of 200 wall-clock minutes.

To the best of our knowledge, this low cost (100 CPU-hours per simulated reactor-relevant plasma) for a full physics integrated modeling workflow is unprecedented, particularly as it includes: quasilinear turbulence modeling with \TGLF, neoclassical transport with \NEO and pedestal modeling with \EPED. We also solve for the internal Grad-Shafranov equilibrium, current diffusion with sawtooth physics, wave heating with \TORIC-\FPPMOD and fast ion physics with \NUBEAM.
All these models are iterated until convergence in their full form, not resorting to surrogate models other than to accelerate the \PORTALS convergence.

\begin{figure}[h!]
    \centering
    \includegraphics[width=1.0\textwidth]{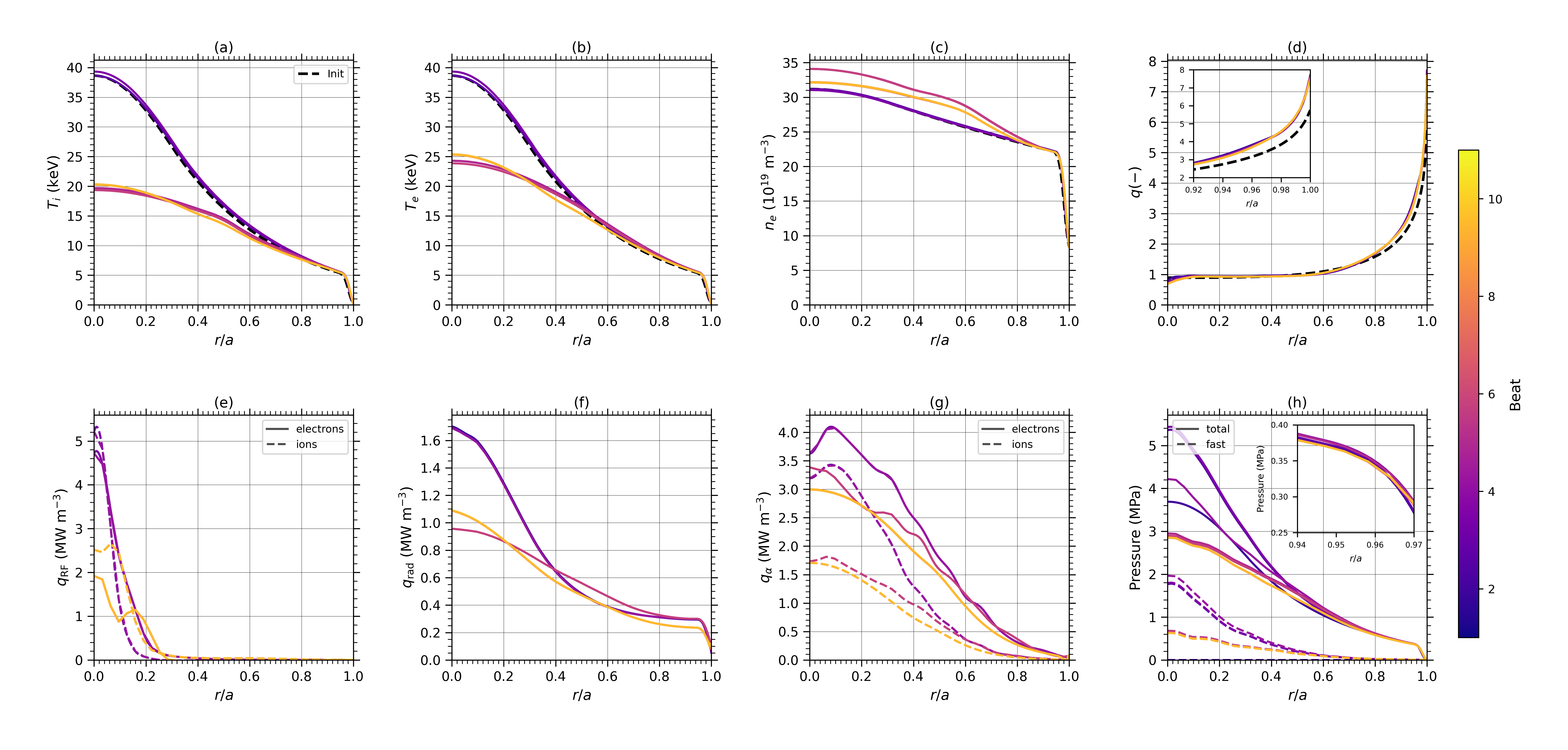}
    \caption{Evolution of specific plasma quantities during the \MAESTRO workflow for the case with $n_{e,\text{ped}}=2.1\times 10^{20}$ m$^{-3}$: (a) ion temperature, (b) electron temperature, (c) electron density, (d) safety factor, (e) absorbed RF power density, (f) radiated power density, (g) alpha power density, and (h) total and fast pressure.
    }
    \label{fig:maestro_beats1}
\end{figure}

Figure~\ref{fig:maestro_beats1} shows the evolution of plasma quantities from beat to beat for the case with $n_{e,\text{ped}}=2.1\times 10^{20}$ m$^{-3}$ as representative of the scan.
Profiles are initialized with a guess of $\beta_N=2.0$, which results in very high temperature profiles (\ref{fig:maestro_beats1}a,b) (and associated radiation (\ref{fig:maestro_beats1}f) and fusion power (\ref{fig:maestro_beats1}g)) that quickly drop in the first \PORTALS beat, where the transport solver finds the flux-matching solution with \TGLF-SAT3 and \NEO at much lower normalized gradients.
The choice of the initial $\beta_N$ and $\nu_{ne}$ guesses can result in a different convergence behavior, and the effect of the initial guess on the final solution will be explored in future work, and may point to the importance of path dependence and full discharge modeling.
The q-profile \ref{fig:maestro_beats1}d evolves from the initial \FreeGS guess to a stationary (pre-sawtooth crash) profile after the first \TRANSP beat, and remains mostly unchanged during the rest of the workflow. Changes in bootstrap current with the evolving pressure profiles only led to minimal changes in current diffusion, given that the first guess of the pedestal pressure from \EPED was already close to the final converged solution, as the $\beta_N$ effect on the pedestal was not significant in these plasma conditions.

\subsection{Efficiency and heuristics}

\begin{figure}[h!]
    \centering
    \includegraphics[width=1.0\textwidth]{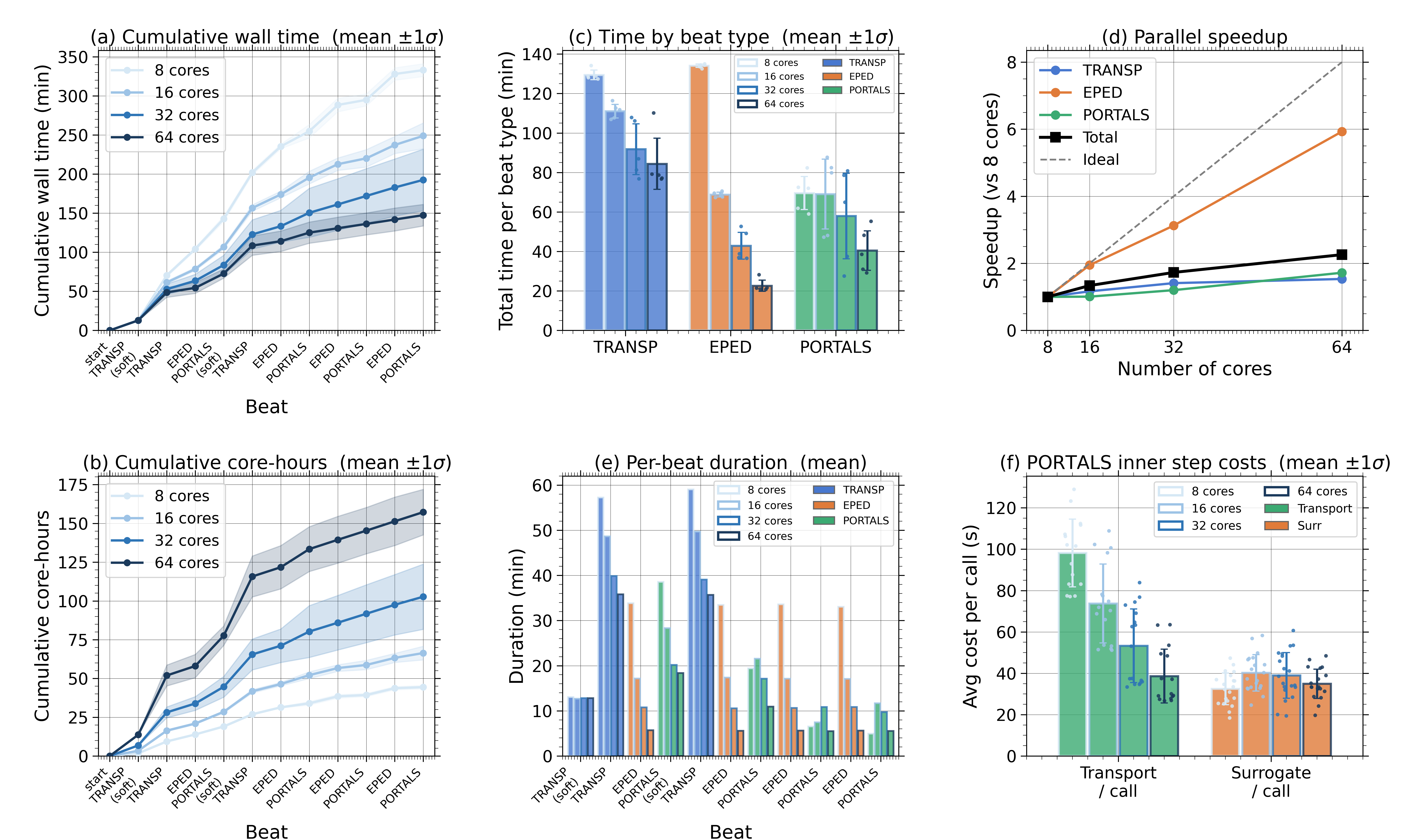}
    \caption{Study of \MAESTRO efficiency (same case as in Figure~\ref{fig:maestro_beats1}) with different random seeds and allocated cores.
    (a) Cumulative wall time and (b) cumulative computational time for the different core allocations, with statistics across 5 different random seeds.
    (c) Time spent per beat type and (e) per individual beat.
    (f) Cost of each inner \PORTALS step, similarly as explored in Figure~\ref{fig:database}.
    (d) Speedup achieved vs the number of cores allocated, with ideal parallel speedup shown as a dashed line.
    }
    \label{fig:maestro_seeds1}
\end{figure}

Figure~\ref{fig:maestro_seeds1} explores the efficiency of the \MAESTRO workflow for the case with $n_{e,\text{ped}}=2.1\times 10^{20}$ m$^{-3}$, exploring different random seeds (which affects the heuristics of \PORTALS) and different core allocations in the AMD EPYC cluster nodes (see Appendix~\ref{sec:Composite} for a description of the computational setup in this machine).
Figures~\ref{fig:maestro_seeds1}a,b,d show clearly that there is no significant speedup when allocating a large number of cores, indicating that there is room of improvement in the parallelization of the workflow.
As shown in figures~\ref{fig:maestro_seeds1}e, the first \TRANSP beat does not benefit from the number of cores, as the current diffusion, internal equilibrium and standard power balance steps are sequential in the code. The parallelization of \NUBEAM and \TORIC in subsequent beats does provide some speedup, but the overall time is still dominated by the sequential steps and I/O operations.
\EPED beats do benefit strongly from the number of cores, showing almost ideal parallel speedup (Figure~\ref{fig:maestro_seeds1}d), while \PORTALS beats show very minimal benefit from the number of cores.
As shown in Figure~\ref{fig:maestro_seeds1}f, the transport model evaluations do get faster with more cores, but the current setup of \PORTALS relies on significant I/O operations for each individual evaluation (particularly for the evaluation of the uncertainty of \TGLF). As expected, the surrogate model operations (training and optimization) do not benefit from the number of cores, as they are clipped to a maximum of 4 threads to avoid over-threading for small operations (see Appendix~\ref{sec:Composite}).

\begin{figure}[h!]
    \centering
    \includegraphics[width=1.0\textwidth]{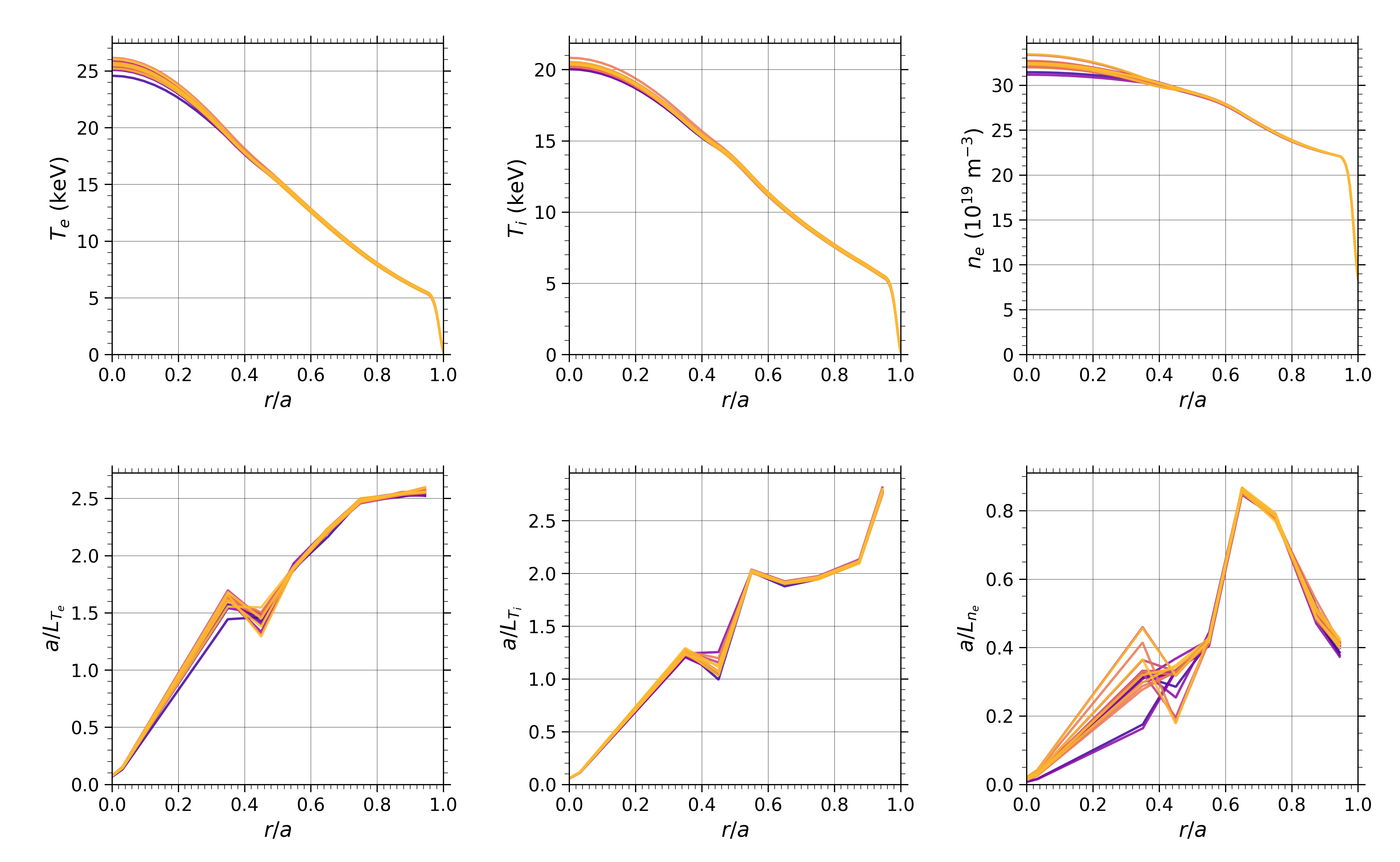}
    \caption{Variation of the final converged profiles for the 20 different random starts (seeds and cores) explored in Figure~\ref{fig:maestro_seeds1}: Ion temperature, electron temperature and electron density profiles, with their corresponding normalized gradients.
    }
    \label{fig:maestro_seeds1_profs}
\end{figure}

The use of surrogate-based optimization via GP models in \PORTALS allows for a significant reduction in the number of transport evaluations required to find the flux-matching solution, but adds a heuristic element to the workflow.
Not only can transport models such as \TGLF-SAT3 exhibit non-unique solutions, but also the technique for early stopping described in Section~\ref{sec:stopping_criteria} that use the uncertainty estimation from Section~\ref{sec:convergence} does not guarantee that the final solution has the exact same flux-matching gradients for all potential random seeds.
Figure~\ref{fig:maestro_seeds1_profs} shows the final converged profiles for the 20 different random starts (seeds and cores) explored in Figure~\ref{fig:maestro_seeds1}, showing that differences can be observed in the final profiles, particularly in the near-axis electron temperature and density profiles.
However, even in such nonlinear burning plasma conditions, the differences in the final predicted fusion power are small, with a standard deviation of $1.7\%$ and maximum variation window of $6\%$.
This is well below the intrinsic uncertainty of the physics models used in the workflow (e.g. pedestal predictions \cite{howard_2026_arc}).

\subsection{On the use of \MAESTRO for core-edge integration studies}

While this manuscript focuses on the presentation of the \MAESTRO workflow and not necessarily on the physics insights obtained from the application to scenario predictions, we note that the modular nature of \MAESTRO allows for the study of core-edge integration in a self-consistent and efficient way.
As described in Section~\ref{sec:MAESTRO}, information flows from the edge to the core through the boundary conditions provided by the pedestal model, and from the core to the edge through the $\beta_N$ input to \EPED.
Furthermore, scrape-off-layer (SOL) and divertor physics can be included in the workflow through the use of the Extended Lengyel model \cite{Body2025Nucl.Fusion_simple} for the calculation of the separatrix temperature and the puffed impurity concentration required for detachment, which can then feed into the pedestal and core plasma through the boundary conditions.

\begin{figure}[h!]
    \centering
    \includegraphics[width=1.0\textwidth]{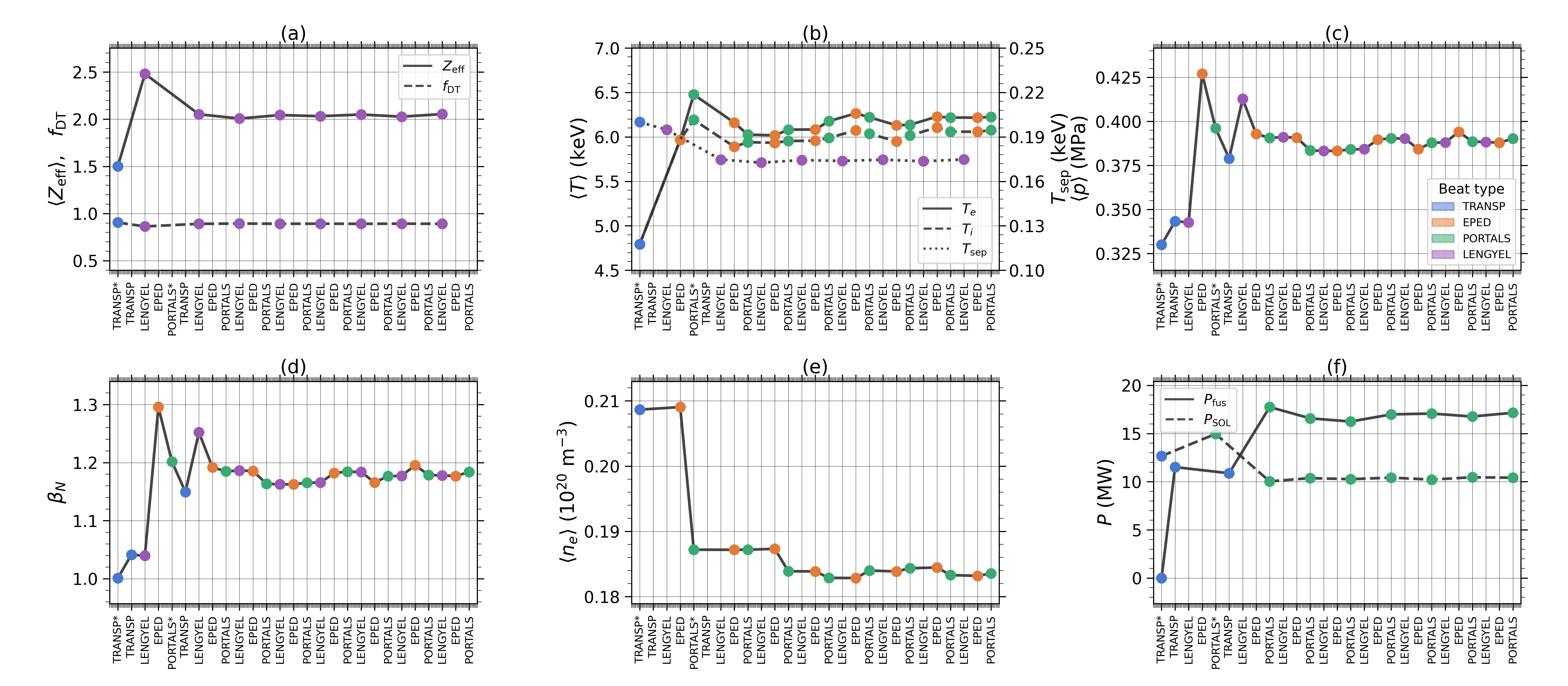}
    \caption{Evolution of quantities during a \MAESTRO simulation. Plots only connect those quantities that are updated in each beat.
    (a) Effective ion charge, $Z_\mathrm{eff}$, and fuel ions concentration, $f_\mathrm{DT}$.
    (b) Volume average ion and electron temperatures, $\langle T_i\rangle$ and $\langle T_e\rangle$, and separatrix temperature, $T_\mathrm{sep}$.
    (c) Volume average total pressure, $\langle p\rangle$, and (d) normalized pressure, $\beta_N$.
    (e) Volume average electron density, $\langle n_e\rangle$.
    (f) Fusion power, $P_\mathrm{fus}$, and scrape-off-layer power, $P_\mathrm{SOL}$.
    }
    \label{fig:maestro_lengyel1}
\end{figure}

Figure~\ref{fig:maestro_lengyel1} shows a predictive simulation of a SPARC \cite{Creely2020J.PlasmaPhys._Overview,Rodriguez-Fernandez2022Nucl.Fusion_Overview} scenario with a reduced magnetic field of $B_T=8.7$T. At this field, the hydrogen minority for ion cyclotron heating resonates off-axis with the nominal antenna frequency of $120$MHz.
The simulation is initialized directly from extended Miller equilibrium parameters \cite{Arbon2020PlasmaPhys.Control.Fusion_Rapidlyconvergent}, with the separatrix temperature and puffed Argon concentration calculated from the Extended Lengyel model, with an enrichment factor \cite{Kallenbach2024Nucl.Fusion_Divertor} of $3.0$, and the pedestal top pressure calculated from \EPED. The core plasma is then evolved with \PORTALS with \TGLF-SAT3 and \NEO until convergence is achieved.
\TRANSP provides the internal equilibrium, current diffusion, wave heating with \TORIC-\FPPMOD, and fast ion physics with \NUBEAM.

The simulation is reasonably well converged after 7 Lengyel-\EPED-\PORTALS iterations, with a total wall-time cost of 8 hours on AMD EPYC cluster nodes with 32 cores.
The high cost of this simulation is mostly due to the fact that \PORTALS cannot currently reutilize surrogate models across the Lengyel-\EPED-\PORTALS iterations, as the impurity concentrations are updated in each iteration and the surrogate models do not include the impurity concentration as an input. This means that each \PORTALS step needs to start from scratch, with no previous knowledge of the transport fluxes, and therefore requiring a larger number of transport evaluations to find the flux-matching solution in each iteration.
This was avoided in the previous application to the ARC scenario (Figure~\ref{fig:maestro_seeds1}) because the pedestal updates from \EPED change the boundary conditions for \PORTALS, whose variation is captured by the gyro-Bohm normalization, the normalized collision rate, and the normalized electron pressure, which are included naturally as inputs in the surrogate models as described in Equation~\ref{eq:vars}.

This points to a clear path for improvement of the \MAESTRO workflow for core-edge integration studies: the inclusion of additional surrogate variables that can capture the changes in transport in between iterations when quantities such as the effective ion charge $Z_\mathrm{eff}$ and fuel ions concentration $f_\mathrm{DT}$ are updated, which would allow for the reutilization of surrogate models across iterations and therefore a significant reduction in the computational cost of the workflow.
Similarly, if one wanted to update the internal equilibrium, including the safety factor and current profiles by running \TRANSP at later stages of the workflow, \PORTALS would need to also include surrogate variables that can capture such changes.
Whether or not all variations in equilibrium and impurity concentrations can be captured by a small number of surrogate variables is an open question, but an interesting avenue for future research is the use of dimensionality reduction techniques to identify the most important surrogate variables that can capture the variations in transport across iterations.

\section{Conclusions}
\label{sec:Conclusions}
This paper has presented a newly developed integrated modeling workflow, \MAESTRO, that couples physics modules to find the self-consistent prediction of tokamak plasmas.
A natural extension of the \PORTALS framework, \MAESTRO enables the coupling of transport, equilibrium, heating, pedestal and edge physics modules in a modular and flexible way, allowing for the use of different solvers for each physics domain depending on the needs of the specific simulation.

The surrogate-based optimization approach at the core of \MAESTRO offers several advantages over traditional time-dependent and Newton-based transport solvers that motivated its development.
The modular, black-box coupling of physics components enables the integration of turbulent transport models of arbitrary fidelity ---from quasilinear eigenvalue solvers to nonlinear gyrokinetics--- without modification to the underlying codes.
The surrogate-based optimization framework minimizes the number of expensive transport flux evaluations required to reach the steady-state solution, which is particularly valuable as higher-fidelity models increase the cost of each evaluation.
By directly targeting the stationary solution as an optimization problem, \MAESTRO bypasses the need to evolve the plasma profiles through physical transients, avoiding the challenge of simulating many confinement times with small time steps imposed by stiffness ---a difficulty that is especially pronounced for particle (density) transport, whose confinement timescale is substantially longer than that of energy.
Additional efficiency gains arise from the re-utilization of previously computed fluxes across \MAESTRO iterations, and from the use of coarse radial grids that avoid the need for heavy radial averaging and can be handled on single compute nodes.

Physics-guided improvements to the \PORTALS framework, such as the positive diffusion constraint and the dynamic simple relaxation technique, together with the tensorization of composite surrogate models, enables the efficient and automated convergence of the transport solver even in the presence of discontinuous transport models.
Large databases of plasma scenarios can be explored in flux-matching conditions with quasilinear transport models without the need to rely on surrogate solutions or low-fidelity physics modeling, and the convergence behavior is robust across a wide range of plasma parameters.

Despite these capabilities, several limitations of the current implementation should be noted.
The most immediate is that only steady-state solutions are possible. Scenarios whose defining physics is inherently time-dependent cannot be fully characterized by a steady-state solution alone.
The physics of q-profile evolution is assumed in this work to result from a long time evolution with fixed kinetic profiles, but the self-consistent consideration of current diffusion and transport evolution could lead to different steady-state solutions.
Capturing such (potentially) trajectory-dependent effects would require time-evolving simulations.

A second limitation is the reliance on heuristic methods, which introduces a degree of path-dependence in the converged solution. As discussed in Section~\ref{sec:Application}, the optimization trajectory ---and therefore the final solution--- can vary with the random seed, since the surrogate-based optimization does not guarantee convergence to a unique flux-matching state.
If the transport physics itself can support multiple steady-state solutions ---e.g., arising from turbulence bifurcations or by the interplay between self-heating and transport--- the optimization of actuator trajectory can determine which solution is found, and currently there is no mechanism to detect or explore the existence of multiple solutions within \PORTALS or \MAESTRO.
In practice the sensitivity of global quantities such as fusion power to the random seed has been found to be small in the cases studied (see Section~\ref{sec:Application} and Ref.~\cite{howard_2026_arc}), but a theoretical guarantee of uniqueness is absent.
We must note, however, that this is not a limitation of the surrogate-based optimization approach itself, but rather of time-independent transport solvers in general.

Finally, the use of coarse radial grids, while advantageous for computational cost, limits the ability to resolve fine-scale profile structures such as internal transport barriers (ITBs), which require fine radial resolution to capture steep gradients accurately.
\PORTALS and \MAESTRO can naturally accommodate finer radial grids, but convergence of the transport solver becomes more difficult and the computational cost increases significantly.
It is worth noting that this challenge is not unique to \PORTALS: time-dependent solvers with finer radial grids face the complementary difficulty that finer spacing reduces the stabilizing effect of radial averaging of turbulent transport coefficients, making convergence harder in ITB regimes, if one wants to still capture the changes in turbulence accross the barrier. A robust and general treatment of ITBs therefore remains an open challenge for integrated transport solvers broadly.

The efficiency of the \MAESTRO workflow has enabled fast iteration with engineering and device designer teams, providing accurate predictions of plasma performance that have directly informed the design of compact fusion power plants. The limitations outlined above define a clear roadmap for future development: extensions to time-dependent transport evolution, adaptive radial grids, and more principled surrogate optimization strategies will further broaden the applicability of \MAESTRO. Together, these advances will strengthen the role of surrogate-accelerated integrated modeling as a predictive and design tool on the path towards fusion energy.

\section*{Acknowledgments}

The \MAESTRO workflow is being developed as part of the MIT Integrated Modeling (MITIM) repository \cite{mitim}, and we encourage interested readers to explore the code and contribute to its open-source development.

The authors would like to thank the members of the MFE Integrated Modeling team at MIT and the Performance and Transport and the SciDAC SMARTS teams for their feedback and support during the development of the \MAESTRO workflow.
We thank J. Candy, E. Belli and G. Staebler for the development of \CGYRO, \NEO and \TGLF, respectively, and the \TRANSP team for the development of \TRANSP.

This work was funded by Commonwealth Fusion Systems under RPP020 and US DoE under grant DE-SC0024399 (SciDAC SMARTS).
Clusters hosted at the Massachusetts Green High Performance Computing Center (MGHPCC) were used to perform the benchmark simulations (MIT-RPP partitions).

OpenAI GPT-5.2 and Claude Sonnet 4.6 were used to enhance parts of the manuscript for clarity and coherence, and to assist in the creation of figures.

\newpage
\begin{appendices}

\section{Generalized transport solver}
\label{app:GeneralizedTransportSolver}

We begin by defining the generic steady-state ($\partial/\partial t \rightarrow 0$) transport equation for channel $Q_c$:
\begin{align}
    \label{eq:transport}
    \langle \mathbf{Q}_c\cdot\nabla r\rangle &=\frac{1}{V'} \int_{0}^{r}\langle S_c \rangle V' dr
\end{align}

Following the terminology used in the transport solver literature, we will refer to the left-hand side of Equation~\ref{eq:transport} as the \textit{transport flux} ($F^{\text{tr}}_c$) and the right-hand side as the \textit{target flux}, with $f^{\text{tar}}_c(r)$ as the \textit{target flux density}.

Discretizing the radial coordinate $r$ of nested flux surfaces into $N_r$ points, and generalizing to potentially several $k_\text{tr}$ sources of transport (e.g. neoclassical and turbulence) and $k_\text{tar}$ targets (e.g. energy exchange, radiation and alpha power), we can write the transport equation for channel $c$ at location $r_j$ as:
\begin{align}
    \label{eq:transport3}
    \sum_{k\in k_\text{tr}} F^{\text{tr},k}_{j,c} &=
    \frac{1}{V'_{j}}
    \int_{0}^{r_j}
    \sum_{k\in k_\text{tar}} f^{\text{tar},k}_c(r)\, V'(r)\, dr
\end{align}

Finding the macroscopic, steady-state plasma solution is then equivalent to solving the set of $N_c\times N_r$ ``channel-radius'' transport equations \ref{eq:transport3}, which can be written as a minimization problem of a scalarized residual function $\xi$:
\begin{align}
    \label{eq:opt1}
    \xi = h^\xi\left(\left\{\sum_{k\in k_\text{tr}} F^{\text{tr},k}_{j,c} - \sum_{k\in k_\text{tar}} h^I_j(f^{\text{tar},k}_{j,c})\right\}_{\forall j,\forall c}\right)
\end{align}
where $h^\xi$ is a nonlinear scalar function that maps the residuals of the individual transport equations (channel and radius) into a scalar (e.g. $L_2$-norm), and the set of $h^I_j$ transformations volume-integrate the target flux densities into the target fluxes at each location $r_j$.
We note that this general formulation can be used to describe the steady-state system for an arbitrary number of (coupled or uncoupled) channels, for any radial discretization, and for any choice of scalarization function.

The goal of a transport solver is then to find the set of input parameters $z_{j,c}$ that minimizes the scalarized residual function $\xi$:
\begin{align}
    \label{eq:opt2}
    &\{z_{j,c}\}^*= \argmin_{z_{j,c}\in [z^L_{j,c},z^U_{j,c}]} \xi(z_{\forall j,\forall c})
\end{align}
where $z_{j,c}$ is, in \PORTALS, the set of local ($r_j$) normalized logarithmic gradients of each channel $c$.
This formulation allows for the separate description of the transport fluxes, $F^{\text{tr},k}_{j,c}$, and the target flux densities, $f^{\text{tar},k}_{j,c}$, which can be provided by separate models or codes, typical of multi-scale plasma simulation frameworks.

In this formulation, the set of local gradients $z_{j,c}$ fully describe the plasma state. The models do not take the full gradient vector directly; instead, nonlinear transformations $h^{\text{tr},k}_{j,c}$ and $h^{\text{tar},k}_{j,c}$ first reduce $z_{\forall j,\forall c}$ to the specific inputs required by each local transport and target model, respectively, so that $F^{\text{tr},k}_{j,c} = F^{\text{tr},k}_{j,c}(h^{\text{tr},k}_{j,c}(z_{\forall j,\forall c}))$ and $f^{\text{tar},k}_{j,c} = f^{\text{tar},k}_{j,c}(h^{\text{tar},k}_{j,c}(z_{\forall j,\forall c}))$.
Details on these transformations are provided extensively in Ref.~\cite{Rodriguez-Fernandez2024Nucl.Fusion_Enhancing}, and involve both radial integration to profile quantities and turbulence-informed transformations to normalized quantities.
Note that, for generalization purposes, input transformations can be different per radial location and per transport channel. In practice, if the transport channels all come from the same simulation, the input transformations can be the same for all channels. But some exceptions may appear when simulating trace impurities \cite{Rodriguez-Fernandez2024Nucl.Fusion_Enhancing}.

Additionally, the output of the models can be provided in normalized quantities (e.g. gyro-Bohm normalization of turbulent transport) and therefore we can include yet another nonlinear transformation to map inputs to their respective output normalizations:
\begin{align}
    \label{eq:outputs1}
    F^{\text{tr},k}_{j,c} &=  g^{\text{tr},k}_{j,c}(z_{\forall j,\forall c})\cdot\widehat{F}^{\text{tr},k}_{j,c}(h^{\text{tr},k}_{j,c}(z_{\forall j,\forall c})) \\
    f^{\text{tar},k}_{j,c} &= g^{\text{tar},k}_{j,c}(z_{\forall j,\forall c})\cdot\widehat{f}^{\text{tar},k}_{j,c}(h^{\text{tar},k}_{j,c}(z_{\forall j,\forall c}))
\end{align}

Putting all this together, the optimization problem reduces to the following description:
\begin{equation}
    \begin{aligned}
        \{z_{j,c}\}^*&= \argmin_{z_{j,c} \in [z^L_{j,c},z^U_{j,c}]} \xi(z_{\forall j,\forall c})\\
        \xi &= h^\xi\left(\left\{\sum_{k\in k_\text{tr}} F^{\text{tr},k}_{j,c} - \sum_{k\in k_\text{tar}} h^I_j(f^{\text{tar},k}_{j,c})\right\}_{\forall j,\forall c}\right) \\
        F^{\text{tr},k}_{j,c} &=  g^{\text{tr},k}_{j,c}(z_{\forall j,\forall c})\cdot\widehat{F}^{\text{tr},k}_{j,c}(h^{\text{tr},k}_{j,c}(z_{\forall j,\forall c})) \\
        f^{\text{tar},k}_{j,c} &= g^{\text{tar},k}_{j,c}(z_{\forall j,\forall c})\cdot\widehat{f}^{\text{tar},k}_{j,c}(h^{\text{tar},k}_{j,c}(z_{\forall j,\forall c}))
    \end{aligned}
    \label{eq:finalopt1}
\end{equation}

To simplify notation, channels $c$ and radial locations $j$ can be included in a set of $m$ models (channels + radii), and $z_{\forall j,\forall c}$ can simply be referred to as $\mathbf{z}$. Setting $h \equiv -h^\xi$ (a sign flip that converts minimization of $\xi$ to maximization, following the \BoTorch \cite{Balandat2020_BoTorch} convention) results in:
\begin{equation}
    \maineq
    \label{eq:finalopt2}
\end{equation}
where the set of $\mathcal{H} = \{h, h^I_m, h^{\text{tr}}_{m,k}, h^{\text{tar}}_{m,k}, g^{\text{tr}}_{m,k}, g^{\text{tar}}_{m,k}\}$ transformations are nonlinear functions, and $\widehat{F}^{\text{tr}}_{m,k}$ and $\widehat{f}^{\text{tar}}_{m,k}$ are the normalized (in their native units) outputs of transport and target models, respectively.

\begin{figure}[h!]
    \centering
    \includegraphics[width=0.6\textwidth]{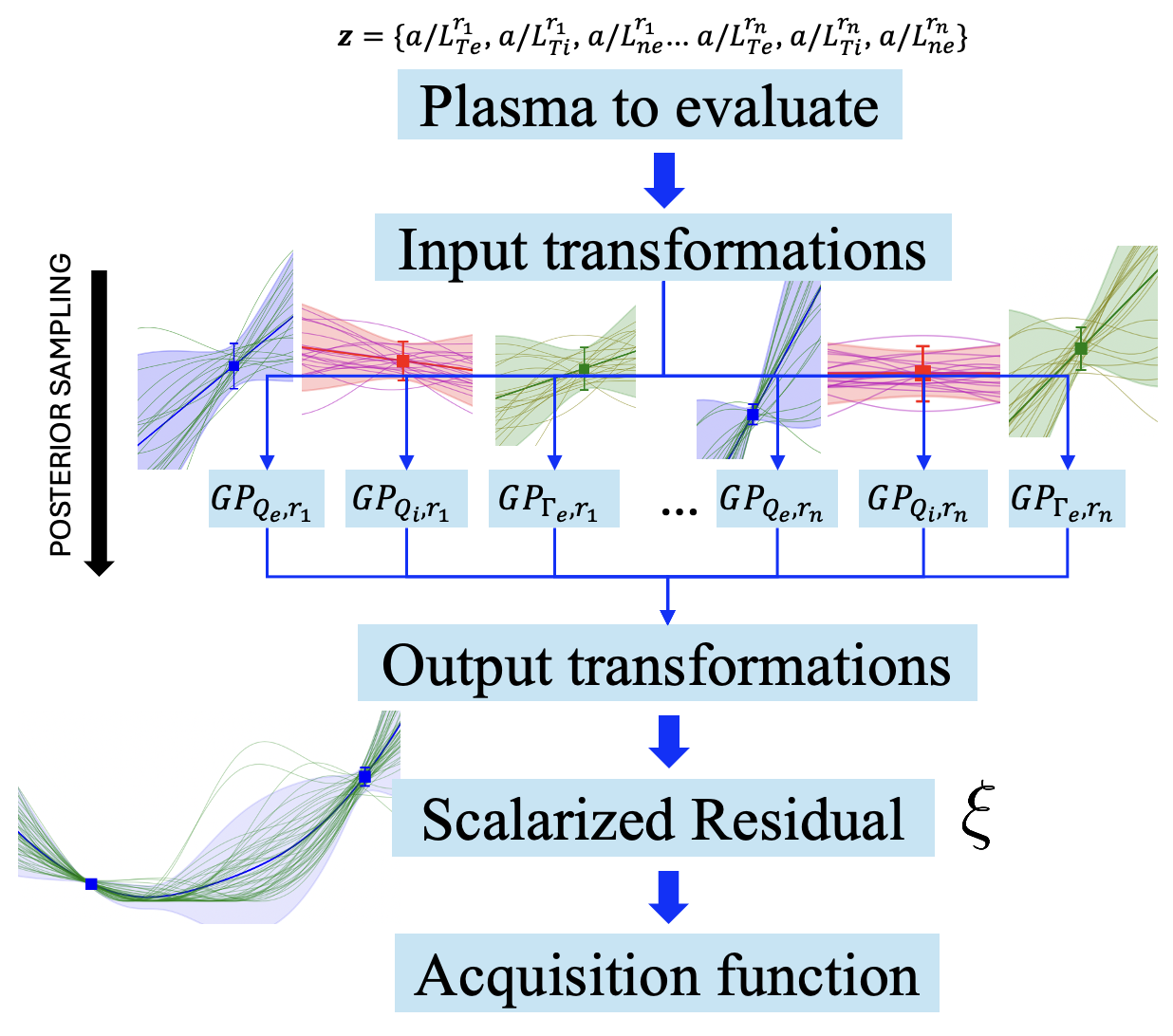}
    \caption{
        Schematic diagram of the composite nature of the surrogate models in \PORTALS.
    }
    \label{fig:GP_diagram}
\end{figure}

Figure~\ref{fig:GP_diagram} shows a visual representation of the composite nature of the surrogate models in \PORTALS. The mapping between the \textit{true} free parameters of the system (the local gradients) and the residual (or acquisition function) is a composition of several transformations. The surrogate models are then built for the normalized transport fluxes and target flux densities, which are then transformed back to physical units and integrated to calculate the residuals of the transport equations.
The GP nature of the surrogate models allows the application of nonlinear transformations to the inputs and outputs of the surrogate models at sample level, which is a powerful tool to define acquisition functions that leverage latest advances in the field of Bayesian optimization.

\subsection{Typical \PORTALS transformations}

In the current implementation of \PORTALS, the transport fluxes are calculated as the sum of neoclassical and turbulent fluxes ($k_{\text{tr}}=2$), and the target fluxes are divided into one analytical components (summation of energy exchange, alpha power and radiation components) and turbulent exchange ($k_{\text{tar}}=2$) in the case of electron and ion energy fluxes.

\begin{itemize}
    \item The scalarization function, $h$, is the negated $L_2$-norm of the residuals, normalized by the total number of channel-radius models $N_m$ (so the metric is independent of problem size and scales as a mean residual magnitude rather than growing with $\sqrt{N_m}$ as the plain $L_2$-norm would):
        \begin{align}
            h(\mathbf{r}) = -\frac{1}{N_m}\cdot \|\mathbf{r}\|_2
        \end{align}
    \item The target flux density integration transformations, $h^I_m$, are defined as the volume integration of the target flux densities, $f^{\text{tar},k}_m$, over the radial coordinate $r$ and are independent of the channel but a function of the radial location:
        \begin{align}
            h^I_m(p) = h^I_j(p) = \int_{V_j} p\, dV = \int_{0}^{r_j} p\, V'(r)\, dr \approx \sum_{i=1}^{j} p_i\, V'_i\, \Delta r_i
        \end{align}
    \item The multi-variable transport-inputs transformations, $h^{\text{tr},k}_m$, are defined such that the input space dimensionality is reduced to the number of inputs required by the local transport model. The transformation is considered the same for the turbulent transport fluxes (all channels), neoclassical transport fluxes  (all channels), and turbulent energy exchange target flux density.
    Here we assume that the local transport simulations depend only on the local values of plasma quantities and the first order derivative only, which results in the reduction to $2\times N_c$ inputs per radial location:
    \begin{align}
        \label{eq:tr_in}
        h^{\text{tr},k}_{m}(\mathbf{z}) = h^{\text{tr}}_{j}(z_{\forall j,\forall c}):
        \quad z_{\forall j,\forall c} &\rightarrow \{z_{j,\forall c}, y_{j,\forall c}\} \rightarrow \{z_{j,\forall c}, \tilde{y}_{j,\forall c}\}
    \end{align}
    where $z_{j,\forall c}$ are the local gradients of the channels, and $y_{j,\forall c}$ are the local values of the channels, obtained via radial integration of the normalized logarithmic gradients \cite{Candy2009Phys.Plasmas_Tokamak}:
    \begin{align}
        y_{j,c} = y_{b,c}\exp\left(\int_{\rho_j}^{\rho_b}z_{c}(\zeta)\,d\zeta\right)
        \approx y_{b,c}\prod_{i=j}^{b-1}\exp\!\left(\frac{z_{i,c}+z_{i+1,c}}{2}\,\Delta\rho_i\right)
    \end{align}
    where $y_{b,c}$ are the boundary values of the channels, $\rho_b$ is the normalized radial coordinate at the boundary, $b$ is the index of the boundary radial location, and $\Delta\rho_i = \rho_{i+1}-\rho_i$.
    
    In Equation~\ref{eq:tr_in}, the values $\tilde{y}_{j,c}$ refer to the physics-guided choice of local plasma parameters that are descriptive of the dimensionless turbulence system and that form a complete basis, $\tilde{y}_{j,c}=g(y_{j,c})$. An example of a standard choice in \PORTALS is, for the simultaneous prediction of $T_e,T_i,n_e,\omega_0,n_Z$:
    \begin{align}
        g(y_j) :
        \quad \{T_e,T_i,n_e,\omega_0,n_Z\}\rightarrow\{\widehat{\nu}_{ei}(T_e,n_e),\frac{T_i}{T_e},\beta_e(T_e,n_e),\frac{\omega_0}{c_s (T_e)},\frac{n_Z}{n_e}\}
        \label{eq:vars}
    \end{align}
    where $\widehat{\nu}_{ei}$ is the normalized electron-ion collision rate, $\beta_e$ is the electron beta, and $c_s$ is the sound speed.
    We note that in the case of trace impurity transport, as described in \cite{Rodriguez-Fernandez2024Nucl.Fusion_Enhancing}, the input transformation will only include impurity gradient in the case of the impurity density channel.

    \item The transport-output transformations, $g^{\text{tr},k}_m$, are designed to unnormalize the gyro-Bohm turbulent transport fluxes, neoclassical transport fluxes, and turbulent energy exchange target flux densities.
    \item The target-input transformations, $h^{\text{tar},k}_m$, for the energy exchange power, alpha heating and radiation power are, given their analytical nature, direct mappings to the output target flux densities, making the GP equal to the identity function. This choice is made for simplicity and to enable the extension of the methodology to more complex targets in the future.
    \item The target-output transformations, $g^{\text{tar},k}_m$, are not used in the current implementation, as the target flux densities are already in real units.
\end{itemize}

\section{Tensorization of composite Gaussian Processes}
\label{sec:Composite}
In previous applications of \PORTALS with quasilinear models it was observed that the optimization of the mean of the posterior distribution acquisition function led to significant overhead cost to find the optimum in the high-dimensional ($N_c\times N_r$, where $N_c$ is the number of channels and $N_r$ is the number of radial locations, thus often $>15$-dimensional) input space. Furthermore, the multi-objective nature of the system (minimization of $N_c\times N_r$ residuals) required careful scalarization or direct multi-objective strategies to avoid poor convergence behavior.
While this is usually not an issue with standard surrogate models (e.g., neural networks), the evaluation of the posterior of Gaussian Processes can become numerically challenging in high-dimensional spaces, particularly in the case of composite surrogate models used here (i.e. separating transport and targets, and local surrogates per radius and channel).
A typical surrogate evaluation in \PORTALS on a problem with $N_r=5$ radii and $N_c=3$ channels with both neoclassical, turbulent and target ($N_m=3$) fluxes requires the evaluation of $N_r\times N_c\times N_m=5\times 3 \times 3 = 45$ GPs, with each GP requiring their own set of input and output space transformations, as described in Appendix~\ref{app:GeneralizedTransportSolver}.

\begin{figure}[h!]
    \centering
    \includegraphics[width=1.0\textwidth]{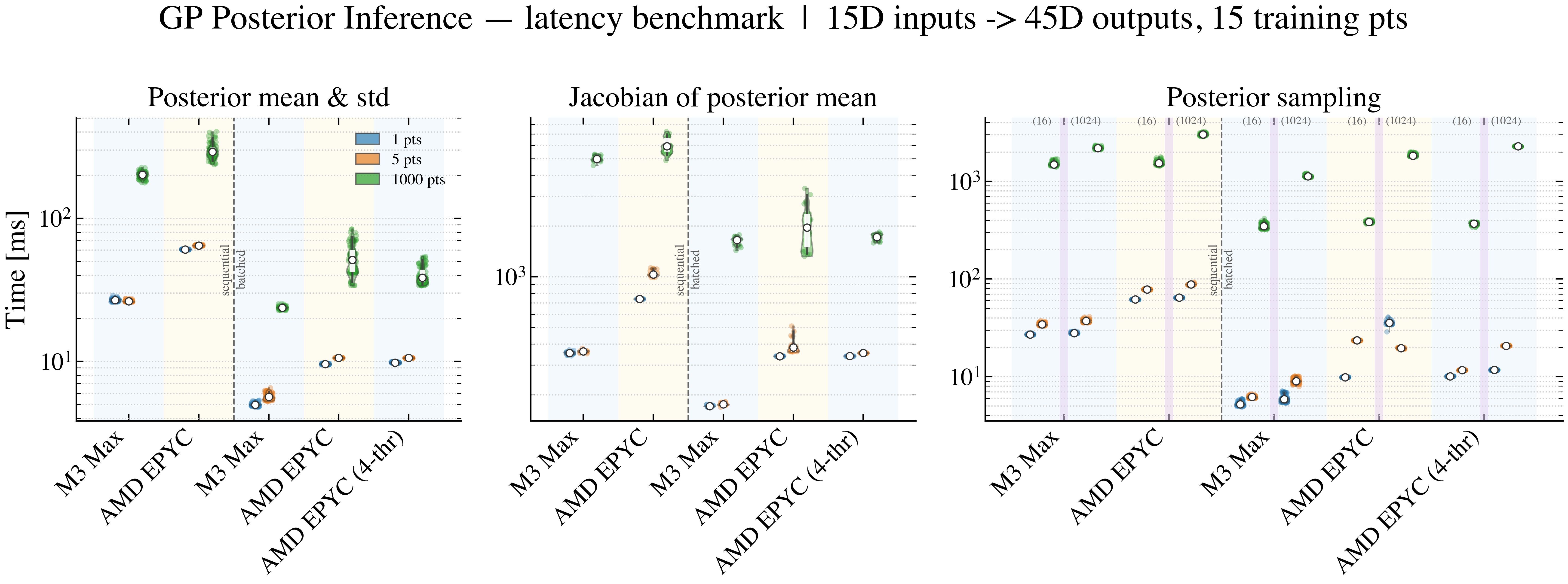}
    \caption{
        Comparison of the computational speed of GP operations in \PORTALS: Evaluating the mean and standard deviation, sampling the posterior distribution ($n=16$ and $n=1024$ samples), and evaluating the Jacobian of the mean.
        Statistics are obtained by averaging over 100 repetitions of each operation.
        Benchmarks were performed on two systems: a laptop (Apple M3 Max, 16-core ARM, 128 GB) and a cluster node (2x AMD EPYC 7543, 64 cores, 512 GB RAM, Rocky Linux 8.10); and for sequential and batched implementations of the GP operations.
    }
    \label{fig:speed_view_comparison}
\end{figure}

Instead of performing $N_r\times N_c\times N_m$ sequential GP evaluations (as in previous versions of \PORTALS), we now construct a composite GP that leverages tensor operations as much as possible (larger matrix multiplications) to reduce dispatch overhead, exploit \GPyTorch \cite{Gardner2021_GPyTorch} efficient machinery and the benefit of multi-threaded systems.
Figure~\ref{fig:speed_view_comparison} shows the significant reduction in evaluation time achieved by using composite GPs and tensor operations, as compared to sequential GP evaluations (previous versions of \PORTALS \cite{Rodriguez-Fernandez2022Nucl.Fusion_Nonlinear,Rodriguez-Fernandez2024Nucl.Fusion_Enhancing}).
This directly validates the batching implementation: sequential calls the GP 45 times (once per output), while batched does it in three grouped passes: one for the boundary locations (with only gradients as input parameters, $z_{BC,\forall c}$), one for the rest of flux evaluations (with gradients and profiles as input parameters, $z_{j,\forall c}$ and $y_{j,\forall c}$), and one for target flux evaluations (different GP kernels).
The 5-8x speedup for the mean and standard deviation (small operations) reflects the reduced dispatch overhead. The Jacobian gap is smaller (2-3x) because the backward pass is compute-bound rather than overhead-bound, so batching has less overhead to eliminate.

Interestingly, the laptop (Apple M3 Max, 16-core ARM, 128 GB) implementation matches or beats the cluster (2x AMD EPYC 7543, 64 cores, 512 GB RAM, Rocky Linux 8.10) implementation for simple operations such as calculating the mean and standard deviation (factor of 2x faster in the laptop), despite having 4x fewer cores. 
This is likely due to the fact that the operations are too small to benefit from many cores (kernel matrices with only 15 training points and 1000 inference points are $[1000 \times 15]$), and the M3's memory subsystem is better for small workloads, particularly when the EPYC cores have to coordinate across 8 separate NUMA domains for what is effectively a cache-sized problem.
Limiting the number of threads in the EPYC implementation to 4 results in an improvement in the speed of these small operations (e.g., 40.6 ms v.s. 52.9 ms for the mean and standard deviation calculation of 1000 points), confirming that over-threading is occurring when each kernel matrix is this small.
For operations with larger matrices (when the $[1000\times1000]$ posterior covariance matrix is built and its Cholesky decomposition is performed), the M3's bandwidth advantage shrinks because the bottleneck shifts from latency to raw throughput, and the EPYC's many cores finally have enough work to justify coordination.

\end{appendices}

\bibliography{References/references.bib,References/references_extra.bib}

@article{Body2025Nucl.Fusion_simple,
	title = {A simple, accurate model for detachment access},
	volume = {65},
	issn = {0029-5515},
	url = {https://doi.org/10.1088/1741-4326/ade4d9},
	doi = {10.1088/1741-4326/ade4d9},
	language = {en},
	number = {8},
	urldate = {2026-03-21},
	journal = {Nuclear Fusion},
	publisher = {IOP Publishing},
	author = {Body, Thomas and Kallenbach, Arne and Eich, Thomas},
	month = jun,
	year = {2025},
	pages = {086002},
}

@article{Kallenbach2024Nucl.Fusion_Divertor,
	title = {Divertor enrichment of recycling impurity species ({He}, {N2}, {Ne}, {Ar}, {Kr}) in {ASDEX} {Upgrade} {H}-modes},
	volume = {64},
	issn = {0029-5515},
	url = {https://doi.org/10.1088/1741-4326/ad3139},
	doi = {10.1088/1741-4326/ad3139},
	language = {en},
	number = {5},
	urldate = {2026-02-03},
	journal = {Nuclear Fusion},
	publisher = {IOP Publishing},
	author = {Kallenbach, A. and Dux, R. and Henderson, S.S. and Tantos, C. and Bernert, M. and Day, C. and McDermott, R.M. and Rohde, V. and Zito, A. and Team, the ASDEX Upgrade},
	month = mar,
	year = {2024},
	pages = {056003},
}

@article{Saltzman2025Nucl.Fusion_Impact,
	title = {Impact of model uncertainty on {SPARC} operating scenario predictions with empirical modeling},
	volume = {66},
	issn = {0029-5515},
	url = {https://doi.org/10.1088/1741-4326/ae2342},
	doi = {10.1088/1741-4326/ae2342},
	language = {en},
	number = {2},
	urldate = {2025-12-19},
	journal = {Nuclear Fusion},
	publisher = {IOP Publishing},
	author = {Saltzman, A. and Rodriguez-Fernandez, P. and Body, T. and Ho, A. and Howard, N.T.},
	month = dec,
	year = {2025},
	pages = {026005},
}

@article{Porcelli1996PlasmaPhys.Control.Fusion_Model,
	title = {Model for the sawtooth period and amplitude},
	volume = {38},
	issn = {0741-3335},
	url = {http://stacks.iop.org/0741-3335/38/i=12/a=010?key=crossref.ccf614f38c2c183df54b01d6933e4f65},
	doi = {10.1088/0741-3335/38/12/010},
	number = {12},
	journal = {Plasma Physics and Controlled Fusion},
	author = {Porcelli, F and Boucher, D and Rosenbluth, M N},
	month = dec,
	year = {1996},
	pages = {2163--2186},
}

@article{Staebler2007Phys.Plasmas_theorybased,
	title = {A theory-based transport model with comprehensive physics},
	volume = {14},
	issn = {1070664X},
	doi = {10.1063/1.2436852},
	number = {5},
	journal = {Physics of Plasmas},
	author = {Staebler, G. M. and Kinsey, J. E. and Waltz, R. E.},
	year = {2007},
	pages = {055909},
}

@article{Staebler2005Phys.Plasmas_GyroLandau,
	title = {Gyro-{Landau} fluid equations for trapped and passing particles},
	volume = {12},
	issn = {1070664X},
	doi = {10.1063/1.2044587},
	number = {10},
	journal = {Physics of Plasmas},
	author = {Staebler, G. M. and Kinsey, J. E. and Waltz, R. E.},
	year = {2005},
	pages = {102508},
}

@article{Bourdelle2005PlasmaPhys.Control.Fusion_Turbulent,
	title = {Turbulent particle transport in magnetized fusion plasma},
	volume = {47},
	issn = {07413335},
	doi = {10.1088/0741-3335/47/5A/023},
	number = {5 A},
	journal = {Plasma Physics and Controlled Fusion},
	author = {Bourdelle, C.},
	year = {2005},
}

@article{Belli2008PlasmaPhys.Control.Fusion_Kinetic,
	title = {Kinetic calculation of neoclassical transport including self-consistent electron and impurity dynamics},
	volume = {50},
	issn = {07413335},
	doi = {10.1088/0741-3335/50/9/095010},
	number = {9},
	journal = {Plasma Physics and Controlled Fusion},
	author = {Belli, E. A. and Candy, J.},
	year = {2008},
	pages = {095010},
}

@article{Candy2009Phys.Plasmas_Tokamak,
	title = {Tokamak profile prediction using direct gyrokinetic and neoclassical simulation},
	volume = {16},
	issn = {1070664X},
	doi = {10.1063/1.3167820},
	number = {6},
	journal = {Physics of Plasmas},
	author = {Candy, J. and Holland, C. and Waltz, R. E. and Fahey, M. R. and Belli, E.},
	year = {2009},
	pages = {060704},
}

@article{LoDestro1994Phys.Plasmas_GradShafranov,
	title = {On the {Grad}-{Shafranov} equation as an eigenvalue problem, with implications for q solvers},
	volume = {1},
	issn = {1070-664X},
	doi = {10.1063/1.870464},
	number = {1},
	journal = {Physics of Plasmas},
	author = {LoDestro, L.L. and {L.D. Pearlstein}},
	year = {1994},
	pages = {90--95},
}

@article{Brambilla1999PlasmaPhys.Control.Fusion_Numerical,
	title = {Numerical simulation of ion cyclotron waves in tokamak plasmas},
	volume = {41},
	number = {1},
	journal = {Plasma Phys. Control. Fusion},
	author = {Brambilla, M},
	year = {1999},
}

@article{Pankin2004Comput.Phys.Commun._tokamak,
	title = {The tokamak {Monte} {Carlo} fast ion module {NUBEAM} in the national transport code collaboration library},
	volume = {159},
	issn = {00104655},
	doi = {10.1016/j.cpc.2003.11.002},
	number = {3},
	journal = {Computer Physics Communications},
	author = {Pankin, Alexei and McCune, Douglas and Andre, Robert and Bateman, Glenn and Kritz, Arnold},
	year = {2004},
	pages = {157--184},
}

@article{Snyder2009Phys.Plasmas_Development,
	title = {Development and validation of a predictive model for the pedestal height},
	volume = {16},
	issn = {1070664X},
	doi = {10.1063/1.3122146},
	number = {5},
	journal = {Physics of Plasmas},
	author = {Snyder, P. B. and Groebner, R. J. and Leonard, A. W. and Osborne, T. H. and Wilson, H. R.},
	year = {2009},
}

@article{Hager2016Phys.Plasmas_Gyrokinetic,
	title = {Gyrokinetic neoclassical study of the bootstrap current in the tokamak edge pedestal with fully non-linear {Coulomb} collisions},
	volume = {23},
	issn = {10897674},
	doi = {10.1063/1.4945615},
	number = {4},
	journal = {Physics of Plasmas},
	author = {Hager, Robert and Chang, C. S.},
	year = {2016},
}

@article{Pereverzev2002IPP-Rep.Max-Planck-Inst.FurPlasmaphys._ASTRA,
	title = {{ASTRA}},
	volume = {IPP 5/98},
	number = {February},
	journal = {IPP-Report (Max-Planck-Institut für Plasmaphysik)},
	author = {Pereverzev, G V and Yushmanov, P N},
	year = {2002},
}

@phdthesis{Hammett1986_Fast,
	title = {Fast ion studies of ion cyclotron heating in the {PLT} tokamak: {Phd} thesises},
	school = {Princeton University},
	author = {Hammett, GW},
	year = {1986},
}

@article{Candy2016J.Comput.Phys._highaccuracy,
	title = {A high-accuracy {Eulerian} gyrokinetic solver for collisional plasmas},
	volume = {324},
	issn = {10902716},
	url = {http://dx.doi.org/10.1016/j.jcp.2016.07.039},
	doi = {10.1016/j.jcp.2016.07.039},
	journal = {Journal of Computational Physics},
	publisher = {Elsevier Inc.},
	author = {Candy, J. and Belli, E. A. and Bravenec, R. V.},
	year = {2016},
	pages = {73--93},
}

@article{Wilson2018Adv.NeuralInf.Process.Syst._Maximizing,
	title = {Maximizing acquisition functions for {Bayesian} optimization},
	volume = {2018-Decem},
	issn = {10495258},
	number = {NeurIPS},
	journal = {Advances in Neural Information Processing Systems},
	author = {Wilson, James T. and Hutter, Frank and Deisenroth, Marc Peter},
	year = {2018},
	note = {arXiv: 1805.10196},
	pages = {9884--9895},
}

@article{Arbon2020PlasmaPhys.Control.Fusion_Rapidlyconvergent,
	title = {Rapidly-convergent flux-surface shape parameterization},
	volume = {63},
	issn = {13616587},
	doi = {10.1088/1361-6587/abc63b},
	number = {1},
	journal = {Plasma Physics and Controlled Fusion},
	author = {Arbon, R. and Candy, J. and Belli, E. A.},
	year = {2020},
	pages = {012001},
}

@article{Ricci2011Phys.Plasmas_Methodology,
	title = {Methodology for turbulence code validation: {Quantification} of simulation-experiment agreement and application to the {TORPEX} experiment},
	volume = {18},
	issn = {1070664X},
	doi = {10.1063/1.3559436},
	number = {3},
	journal = {Physics of Plasmas},
	author = {Ricci, Paolo and Theiler, C. and Fasoli, A. and Furno, I. and Gustafson, K. and Iraji, D. and Loizu, J.},
	year = {2011},
}

@article{Creely2020J.PlasmaPhys._Overview,
	title = {Overview of the {SPARC} tokamak},
	volume = {86},
	copyright = {All rights reserved},
	issn = {0022-3778},
	url = {https://www.cambridge.org/core/journals/journal-of-plasma-physics/article/overview-of-the-sparc-tokamak/DD3C44ECD26F5EACC554811764EF9FF0},
	doi = {10.1017/S0022377820001257},
	number = {5},
	urldate = {2021-07-12},
	journal = {Journal of Plasma Physics},
	publisher = {Cambridge University Press},
	author = {Creely, A. J. and Greenwald, M. J. and Ballinger, S. B. and Brunner, D. and Canik, J. and Doody, J. and Fülöp, T. and Garnier, D. T. and Granetz, R. and Gray, T. K. and Holland, C. and Howard, N. T. and Hughes, J. W. and Irby, J. H. and Izzo, V. A. and Kramer, G. J. and Kuang, A. Q. and LaBombard, B. and Lin, Y. and Lipschultz, B. and Logan, N. C. and Lore, J. D. and Marmar, E. S. and Montes, K. and Mumgaard, R. T. and Paz-Soldan, C. and Rea, C. and Reinke, M. L. and Rodriguez-Fernandez, P. and Särkimäki, K. and Sciortino, F. and Scott, S. D. and Snicker, A. and Snyder, P. B. and Sorbom, B. N. and Sweeney, R. and Tinguely, R. A. and Tolman, E. A. and Umansky, M. and Vallhagen, O. and Varje, J. and Whyte, D. G. and Wright, J. C. and Wukitch, S. J. and Zhu, J. and {SPARC Team}},
	year = {2020},
	pages = {865860502},
}

@article{Rodriguez-Fernandez2020J.PlasmaPhys._Predictions,
	title = {Predictions of core plasma performance for the {SPARC} tokamak},
	volume = {86},
	copyright = {All rights reserved},
	issn = {0022-3778},
	url = {https://www.cambridge.org/core/journals/journal-of-plasma-physics/article/predictions-of-core-plasma-performance-for-the-sparc-tokamak/3B88CC267B001A283138318FEE3CDC3D},
	doi = {10.1017/S0022377820001075},
	number = {5},
	urldate = {2021-07-12},
	journal = {Journal of Plasma Physics},
	publisher = {Cambridge University Press},
	author = {Rodriguez-Fernandez, P. and Howard, N. T. and Greenwald, M. J. and Creely, A. J. and Hughes, J. W. and Wright, J. C. and Holland, C. and Lin, Y. and Sciortino, F. and Team, the SPARC},
	year = {2020},
	pages = {865860503},
}

@article{Breslau2018Comput.Softw.USDOEOff.Sci.SCFusionEnergySci.FES_TRANSP,
	title = {{TRANSP}},
	doi = {10.11578/DC.20180627.4},
	urldate = {2022-01-04},
	journal = {Computer software. USDOE Office of Science (SC), Fusion Energy Sciences (FES)},
	author = {Breslau, Joshua and Gorelenkova, Marina and Poli, Francesca and Sachdev, Jai and Pankin, Alexei and Perumpilly, Gopan and Yuan, Xingqiu and Glant, Laszlo},
	month = jun,
	year = {2018},
}

@article{Rodriguez-Fernandez2022Nucl.Fusion_Overview,
	title = {Overview of the {SPARC} physics basis towards the exploration of burning-plasma regimes in high-field, compact tokamaks},
	volume = {62},
	copyright = {All rights reserved},
	issn = {0029-5515},
	url = {https://iopscience.iop.org/article/10.1088/1741-4326/ac1654},
	doi = {10.1088/1741-4326/AC1654},
	number = {4},
	urldate = {2022-03-01},
	journal = {Nuclear Fusion},
	publisher = {IOP Publishing},
	author = {Rodriguez-Fernandez, P. and Creely, A.J. and Greenwald, M.J. and Brunner, D. and Ballinger, S.B. and Chrobak, C.P. and Garnier, D.T. and Granetz, R. and Hartwig, Z.S. and Howard, N.T. and Hughes, J.W. and Irby, J.H. and Izzo, V.A. and Kuang, A.Q. and Lin, Y. and Marmar, E.S. and Mumgaard, R.T. and Rea, C. and Reinke, M.L. and Riccardo, V. and Rice, J.E. and Scott, S.D. and Sorbom, B.N. and Stillerman, J.A. and Sweeney, R. and Tinguely, R.A. and Whyte, D.G. and Wright, J.C. and Yuryev, D.V.},
	month = mar,
	year = {2022},
	pages = {042003},
}

@article{Lyons2023PhysicsofPlasmas_Flexible,
	title = {Flexible, integrated modeling of tokamak stability, transport, equilibrium, and pedestal physics},
	volume = {30},
	issn = {1070-664X},
	url = {https://doi.org/10.1063/5.0156877},
	doi = {10.1063/5.0156877},
	number = {9},
	urldate = {2025-03-18},
	journal = {Physics of Plasmas},
	author = {Lyons, B. C. and McClenaghan, J. and Slendebroek, T. and Meneghini, O. and Neiser, T. F. and Smith, S. P. and Weisberg, D. B. and Belli, E. A. and Candy, J. and Hanson, J. M. and Lao, L. L. and Logan, N. C. and Saarelma, S. and Sauter, O. and Snyder, P. B. and Staebler, G. M. and Thome, K. E. and Turnbull, A. D.},
	month = sep,
	year = {2023},
	pages = {092510},
}

@misc{Meneghini2024_FUSE,
	title = {{FUSE} ({Fusion} {Synthesis} {Engine}): {A} {Next} {Generation} {Framework} for {Integrated} {Design} of {Fusion} {Pilot} {Plants}},
	shorttitle = {{FUSE} ({Fusion} {Synthesis} {Engine})},
	url = {http://arxiv.org/abs/2409.05894},
	doi = {10.48550/arXiv.2409.05894},
	urldate = {2024-11-21},
	publisher = {arXiv},
	author = {Meneghini, O. and Slendebroek, T. and Lyons, B. C. and McLaughlin, K. and McClenaghan, J. and Stagner, L. and Harvey, J. and Neiser, T. F. and Ghiozzi, A. and Dose, G. and Guterl, J. and Zalzali, A. and Cote, T. and Shi, N. and Weisberg, D. and Smith, S. P. and Grierson, B. A. and Candy, J.},
	month = sep,
	year = {2024},
	note = {arXiv:2409.05894},
}

@misc{Citrin2024_TORAX,
	title = {{TORAX}: {A} {Fast} and {Differentiable} {Tokamak} {Transport} {Simulator} in {JAX}},
	shorttitle = {{TORAX}},
	url = {http://arxiv.org/abs/2406.06718},
	doi = {10.48550/arXiv.2406.06718},
	urldate = {2024-06-17},
	publisher = {arXiv},
	author = {Citrin, Jonathan and Goodfellow, Ian and Raju, Akhil and Chen, Jeremy and Degrave, Jonas and Donner, Craig and Felici, Federico and Hamel, Philippe and Huber, Andrea and Nikulin, Dmitry and Pfau, David and Tracey, Brendan and Riedmiller, Martin and Kohli, Pushmeet},
	month = jun,
	year = {2024},
	note = {arXiv:2406.06718 [physics]},
}

@article{Rodriguez-Fernandez2024Nucl.Fusion_Enhancing,
	title = {Enhancing predictive capabilities in fusion burning plasmas through surrogate-based optimization in core transport solvers},
	volume = {64},
	copyright = {All rights reserved},
	issn = {0029-5515},
	url = {https://dx.doi.org/10.1088/1741-4326/ad4b3d},
	doi = {10.1088/1741-4326/ad4b3d},
	language = {en},
	number = {7},
	urldate = {2024-06-05},
	journal = {Nuclear Fusion},
	publisher = {IOP Publishing},
	author = {Rodriguez-Fernandez, P. and Howard, N. T. and Saltzman, A. and Kantamneni, S. and Candy, J. and Holland, C. and Balandat, M. and Ament, S. and White, A. E.},
	month = jun,
	year = {2024},
	pages = {076034},
}

@article{Rodriguez-Fernandez2024PhysicsofPlasmas_Core,
	title = {Core performance predictions in projected {SPARC} first-campaign plasmas with nonlinear {CGYRO}},
	volume = {31},
	copyright = {All rights reserved},
	issn = {1070-664X},
	url = {https://doi.org/10.1063/5.0209752},
	doi = {10.1063/5.0209752},
	number = {6},
	urldate = {2024-06-03},
	journal = {Physics of Plasmas},
	author = {Rodriguez-Fernandez, P. and Howard, N. T. and Saltzman, A. and Shoji, L. and Body, T. and Battaglia, D. J. and Hughes, J. W. and Candy, J. and Staebler, G. M. and Creely, A. J.},
	month = jun,
	year = {2024},
	pages = {062501},
}

@misc{Astudillo2019_Bayesian,
	title = {Bayesian {Optimization} of {Composite} {Functions}},
	url = {http://arxiv.org/abs/1906.01537},
	doi = {10.48550/arXiv.1906.01537},
	urldate = {2024-05-21},
	publisher = {arXiv},
	author = {Astudillo, Raul and Frazier, Peter I.},
	month = jun,
	year = {2019},
	note = {arXiv:1906.01537 [cs, math, stat]},
}

@misc{Gardner2021_GPyTorch,
	title = {{GPyTorch}: {Blackbox} {Matrix}-{Matrix} {Gaussian} {Process} {Inference} with {GPU} {Acceleration}},
	shorttitle = {{GPyTorch}},
	url = {http://arxiv.org/abs/1809.11165},
	doi = {10.48550/arXiv.1809.11165},
	urldate = {2024-05-21},
	publisher = {arXiv},
	author = {Gardner, Jacob R. and Pleiss, Geoff and Bindel, David and Weinberger, Kilian Q. and Wilson, Andrew Gordon},
	month = jun,
	year = {2021},
	note = {arXiv:1809.11165 [cs, stat]},
}

@misc{Balandat2020_BoTorch,
	title = {{BoTorch}: {A} {Framework} for {Efficient} {Monte}-{Carlo} {Bayesian} {Optimization}},
	shorttitle = {{BoTorch}},
	url = {http://arxiv.org/abs/1910.06403},
	doi = {10.48550/arXiv.1910.06403},
	urldate = {2023-12-19},
	publisher = {arXiv},
	author = {Balandat, Maximilian and Karrer, Brian and Jiang, Daniel R. and Daulton, Samuel and Letham, Benjamin and Wilson, Andrew Gordon and Bakshy, Eytan},
	month = dec,
	year = {2020},
	note = {arXiv:1910.06403 [cs, math, stat]},
}

@article{MolinaCabrera2023PhysicsofPlasmas_Isotope,
	title = {Isotope effects on energy transport in the core of {ASDEX}-{Upgrade} tokamak plasmas: {Turbulence} measurements and model validation},
	volume = {30},
	copyright = {All rights reserved},
	issn = {1070-664X},
	shorttitle = {Isotope effects on energy transport in the core of {ASDEX}-{Upgrade} tokamak plasmas},
	url = {https://doi.org/10.1063/5.0143416},
	doi = {10.1063/5.0143416},
	number = {8},
	urldate = {2023-08-22},
	journal = {Physics of Plasmas},
	author = {Molina Cabrera, P. A. and Rodriguez-Fernandez, P. and Görler, T. and Bergmann, M. and Höfler, K. and Denk, S. S. and Bielajew, R. and Conway, G. D. and Yoo, C. and White, A. E. and {ASDEX Upgrade Team}},
	month = aug,
	year = {2023},
	pages = {082304},
}

@misc{Mandell2022_GX,
	title = {{GX}: a {GPU}-native gyrokinetic turbulence code for tokamak and stellarator design},
	shorttitle = {{GX}},
	url = {https://arxiv.org/abs/2209.06731v3},
	doi = {10.48550/arXiv.2209.06731},
	language = {en},
	urldate = {2023-02-28},
	author = {Mandell, N. R. and Dorland, W. and Abel, I. and Gaur, R. and Kim, P. and Martin, M. and Qian, T.},
	month = sep,
	year = {2022},
}

@article{Siena2023Nucl.Fusion_Predictions,
	title = {Predictions of improved confinement in {SPARC} via energetic particle turbulence stabilization},
	volume = {63},
	copyright = {All rights reserved},
	issn = {0029-5515},
	url = {https://dx.doi.org/10.1088/1741-4326/acb1c7},
	doi = {10.1088/1741-4326/acb1c7},
	language = {en},
	number = {3},
	urldate = {2023-02-01},
	journal = {Nuclear Fusion},
	publisher = {IOP Publishing},
	author = {Siena, A. Di and Rodriguez-Fernandez, P. and Howard, N. T. and Navarro, A. Bañón and Bilato, R. and Görler, T. and Poli, E. and Merlo, G. and Wright, J. and Greenwald, M. and Jenko, F.},
	month = jan,
	year = {2023},
	pages = {036003},
}

@article{Dudding2022Nucl.Fusion_new,
	title = {A new quasilinear saturation rule for tokamak turbulence with application to the isotope scaling of transport},
	volume = {62},
	issn = {0029-5515},
	url = {https://doi.org/10.1088/1741-4326/ac7a4d},
	doi = {10.1088/1741-4326/ac7a4d},
	language = {en},
	number = {9},
	urldate = {2022-07-14},
	journal = {Nuclear Fusion},
	publisher = {IOP Publishing},
	author = {Dudding, H. G. and Casson, F. J. and Dickinson, D. and Patel, B. S. and Roach, C. M. and Belli, E. A. and Staebler, G. M.},
	month = jul,
	year = {2022},
	pages = {096005},
}

@article{Rodriguez-Fernandez2022Nucl.Fusion_Nonlinear,
	title = {Nonlinear gyrokinetic predictions of {SPARC} burning plasma profiles enabled by surrogate modeling},
	volume = {62},
	copyright = {All rights reserved},
	issn = {0029-5515},
	url = {https://iopscience.iop.org/article/10.1088/1741-4326/ac64b2},
	doi = {10.1088/1741-4326/AC64B2},
	number = {7},
	urldate = {2022-05-16},
	journal = {Nuclear Fusion},
	publisher = {IOP Publishing},
	author = {Rodriguez-Fernandez, P. and Howard, N.T. and Candy, J.},
	month = may,
	year = {2022},
	pages = {076036},
}

@misc{mitim,
  title        = {MITIM: a toolbox for modeling tasks in plasma physics and fusion energy},
  author       = {Rodriguez-Fernandez, Pablo},
  year         = {2024},
  note         = {Version 1.1},
  howpublished = {\url{https://github.com/pabloprf/MITIM-fusion}},
  url          = {https://mitim-fusion.readthedocs.io/en/latest/}
}

@misc{freegs,
  title        = {{FreeGS}: Free boundary Grad-Shafranov solver},
  author       = {Dudson, Ben},
  year         = {2024},
  howpublished = {\url{https://freegs.readthedocs.io/en/latest/}},
  url          = {https://freegs.readthedocs.io/en/latest/}
}

@article{howard_2026_arc,
  title = {Performance and Transport in the {ARC} Tokamak},
  journal = {Journal of Plasma Physics (accepted)},
  author = {Howard, N. T. and Rodriguez-Fernandez, P. and Hall, J. and Muraca, M. and Saltzman, A. and Ho, A. and Hillesheim, J. C. and Creely, A.},
  year = {2026},
}

@article{hillesheim_2026_arc_physics_basis,
  title = {Overview of the physics basis for the {ARC} fusion power plant},
  journal = {Journal of Plasma Physics (accepted)},
  author = {Hillesheim, J. C. and Creely, A. J. and Eich, T. H. and Howard, N. T. and Leuthold, N. and Sweeney, R. and LeViness, A. and Nelson, A. O. and Nichols, L. and Tinguely, R. A. and Usoltseva, M. and Battaglia, D. and Body, T. A. J. and Hansen, C. and Logan, N. C. and Mumgaard, R. T. and Rodriguez-Fernandez, P. and Snyder, P. B. and Sorbom, B. N. and Wright, J. C.},
  year = {2026},
}

\end{document}